\pdfoutput=1
\documentclass[11pts]{article}

\usepackage{enumerate}
\usepackage{hyperref}
\usepackage[utf8]{inputenc}
\usepackage[T1]{fontenc}
\usepackage[english]{babel}
\usepackage{url}
\usepackage[squaren, Gray, cdot]{SIunits}
\usepackage{array}
\usepackage{caption}
\usepackage{layout}
\usepackage{csquotes}
\usepackage[top=3.2cm,bottom=3.2cm,left=3.2cm,right=3.2cm]{geometry}

\usepackage{amsthm}
\usepackage{amsmath}
\usepackage{amssymb}
\usepackage{thm-restate}

\newtheorem{theorem}{Theorem}
\newtheorem{definition}[theorem]{Definition}
\newtheorem{lemma}[theorem]{Lemma}

\newtheorem{claim}[theorem]{Claim}
\newtheorem{remark}[theorem]{Remark}
\newtheorem{corollary}[theorem]{Corollary}

\usepackage{listings}

\usepackage{graphicx}
\usepackage{subfigure}

\usepackage{bbm}
\usepackage{stmaryrd}
\usepackage{tikz}

\usepackage{algorithm}
\usepackage[noend]{algpseudocode}

\allowdisplaybreaks

\title{Matroid-Constrained Vertex Cover}

\date{}

\author{Chien-Chung Huang\\
\textit{CNRS, DI ENS, École normale supérieure, Université PSL, France}
\and François Sellier\\
\textit{Université Paris Cité, CNRS, IRIF, Paris, France}\\
\textit{Mines Paris, Université PSL, Paris, France}}

\begin{document}

\maketitle
\begin{abstract}
    In this paper, we introduce the problem of \emph{Matroid-Constrained Vertex Cover}: given a graph with weights on the edges and a matroid imposed on the vertices, our problem is to choose a subset of vertices that is independent in the matroid, with the objective of maximizing the total weight of covered edges. This problem is a generalization of the much studied \textsc{max $k$-vertex cover} problem, in which the matroid is the simple uniform matroid, and it is also a special case of the problem of maximizing a monotone submodular function under a matroid constraint.
    
    In the first part of this work, we give a Fixed-Parameter Tractable Approximation Scheme (FPT-AS) when the given matroid is a partition matroid, a laminar matroid, or a transversal matroid. Precisely, if $k$ is the rank of the matroid, we obtain $(1 - \varepsilon)$ approximation using  $\left(\frac{1}{\varepsilon}\right)^{O(k)}n^{O(1)}$ time for partition and laminar matroids and using $\left(\frac{1}{\varepsilon}+k\right)^{O(k)}n^{O(1)}$ time for transversal matroids. This extends a result of Manurangsi for uniform matroids~\cite{manurangsi2018note}. We also show that these ideas can be applied in the context of (single-pass) streaming algorithms. Besides, our FPT-AS introduces a new technique based on matroid union, which may be of independent interest in extremal combinatorics.
    
    In the second part, we consider general matroids. We propose a simple local search algorithm that guarantees $\frac{2}{3} \approx 0.66$ approximation. For the more general problem where two matroids are imposed on the vertices and a feasible solution must be a common independent set, we show that a local search algorithm gives a $\frac{2}{3} \cdot \left(1 - \frac{1}{p+1}\right)$ approximation in $n^{O(p)}$ time, for any integer $p$. We also provide some evidence to show that with the constraint of one or two matroids, the approximation ratio of $2/3$ is likely the best possible, using the currently known techniques of local search.
\end{abstract}

\paragraph*{Keywords} Vertex cover, matroid, kernel, local search

\paragraph*{Acknowledgements} This work was funded by the grants ANR-19-CE48-0016 and ANR-18-CE40-0025-01 from the French National Research Agency (ANR).

\paragraph*{Related version} The first part of the present work has been presented at SWAT 2022~\cite{HuangSWAT2022}.

\section{Introduction}

    \paragraph{The Matroid-Constrained Vertex Cover Problem} Let $G = (V, E)$ be a graph. A weight $w(e)$ is associated with each edge $e \in E$. By convention, we set $n = |V|$ and $m = |E|$. For a vertex $v \in V$ we define $\delta(v)$ the set of edges that are incident to $v$. 
    The \emph{degree} of a vertex $v \in V$, denoted $\deg(v)$, is the size of $\delta(v)$, and we define the \emph{weighted degree} of a vertex $v \in V$ as the sum $\deg_w(v) = \sum_{e \in \delta(v)} w(e)$. For two sets of vertices $S, T \subseteq V$ in a graph $G$, we denote $E_G(S, T) = \sum_{e \in E, e \cap S \neq \emptyset, e \cap T \neq \emptyset}w(e)$ the sum of the weights of the edges that have at least one endpoint in $S$ and at least one endpoint in $T$. Then $E_G(S, S)$, abbreviated $E_G(S)$, denotes the sum of the weights of the edges that are covered by $S$ (\emph{i.e.} having at least one of its endpoints in $S$).
    
    Let $\mathcal{M} = (V, \mathcal{I})$ be a matroid on the ground set $V$. Recall that $\mathcal{M} = (V, \mathcal{I})$ is a matroid if the following three conditions hold: (1) $\emptyset \in \mathcal{I}$, (2) if $X\subseteq Y \in \mathcal{I}$, then $X\in \mathcal{I}$, and (3) if $X, Y \in \mathcal{I}, |Y| > |X|$, there exists an element $e \in Y \backslash X$ so that $X \cup \{e\} \in \mathcal{I}$. The sets in $\mathcal{I}$ are the \emph{independent sets} and the \emph{rank} $k$ of the matroid $\mathcal{M}$ is defined as $\max_{X \in \mathcal{I}}|X|$. For more details about matroids, we refer the reader to~\cite{Schrijver2003}. In this paper, given a set $S \subseteq V$ and $v \in V$, we will denote $S \cup \{v\}$ by $S + v$ and $S \backslash \{v\}$ by $S - v$ for conciseness. 
    
    The problem that we consider in this paper is to choose an independent 
    set of vertices $S \in \mathcal{I}$, with the objective of maximizing 
    $E_G(S)$, namely, the total weight of the edges covered by $S$.
    
    Let us put our problem in a larger picture. When the given matroid $\mathcal{M}$ is a uniform matroid (see below for a formal definition), 
    our problem reduces to the \textsc{max $k$-vertex-cover} problem, where we want to 
    choose $k$ arbitrary vertices so as to maximize the total weight of covered edges. This is a classical problem with a long history: the greedy heuristic is known to give $1-1/e$ approximation as shown by Hochbaum and Pathria~\cite{Hochbaum1998}. 
    Ageev and Sviridenko~\cite{AgeevS99} propose an LP-based approach and the technique of pipage rounding to obtain $3/4$ approximation. Using SDP, Feige and Langberg~\cite{FEIGE2001174} improve this ratio to $3/4+\delta$ for some small constant $\delta>0$. 
    The current best approximation ratio is $0.92$, achieved by Manurangsi~\cite{manurangsi2018note}. For some special cases of the problem, different ratios are also obtained, \emph{e.g.}, see~\cite{BEPS18,HYZZ02,HYZ02}. On the hardness side, to our knowledge, the best inapproximability ratio is due to Austrin and Stankovic~\cite{DBLP:conf/approx/AustrinS19}, which is $0.929$. 
    
    The \textsc{max $k$-vertex-cover} has also been studied through the lens of 
    fixed-parameterized-tractability. Guo et al.~\cite{Guo2005} show 
    the problem to be $W[1]$-hard with $k$ as parameter, thus showing the unlikelihood of getting an exact solution in FPT time. Nonetheless,  Marx~\cite{Marx2008} shows that it is possible to get a near-optimal solution in FPT time. Precisely, he gives an FPT approximation scheme (FPT-AS), that delivers a $(1-\varepsilon)$-approximate solution 
    in $(k/\varepsilon)^{O(k^3/\varepsilon)}n^{O(1)}$ time. This running time is later improved by Gupta et al.~\cite{GuptaLL18} and Manurangsi~\cite{manurangsi2018note}. 
    
    Here we recall the definition of an FPT-AS~\cite{Marx2008}:
    \begin{definition}
        Given a parameter function $\kappa$ associating a natural number to each instance $x \in I$ of a given problem, a Fixed-Parameter Tractable Approximation Scheme (FPT-AS) is an algorithm that can provide a $(1-\varepsilon)$ approximation in $f(\varepsilon, \kappa(x)) \cdot |x|^{O(1)}$ time.
    \end{definition}
    In our case, the instances are made of a graph and a matroid, and the parameter of an instance is the rank $k$ of its matroid.
    
    Regarding the more general case of an arbitrary matroid of rank $k$, one can obtain $3/4$ approximation in polynomial time by combining known techniques.\footnote{Ageev and Sviridenko~\cite{AgeevS99} show that, for the case of a uniform matroid, the optimal fractional solution $x^*$ of the LP has at least $3/4$ of the optimal value. They then use the pipage rounding to transform it into an integral solution with value no less than $x^*$. The same LP approach can be generalized for arbitrary matroids. The optimal fractional solution can be obtained by the Ellipsoid algorithm: even though the linear program to describe the independent sets of an arbitrary matroid may use exponentially many constraints, we can design a separation oracle using an algorithm of Cunningham~\cite{Cunningham1984}. What remains is just the pipage rounding with a general matroid---this is already known to be do-able by Calinescu et al.~\cite{Calinescu2011}. We thank Pasin Manurangsi for communicating to us this method.} This is also a special case of maximizing a submodular function (more precisely, a coverage function) under matroid constraint, for which a $1 - 1/e$ approximation can be achieved in polynomial time~\cite{Calinescu2007,Filmus2012}. 
    In this work, we try to do better than this ratio for some special cases of matroids, in the context of fixed-parameter algorithms. We also show that the ideas developed here can be applied in the streaming setting~\cite{Mut2005}.
    In streaming, maximizing a submodular function under a general matroid constraint has received much attention recently~\cite{Chakrabarti2014,ChekuriGQ15,FeldmanLNSZ22}. Then we also show how local search can be used to obtain simple algorithms for this problem and what are the limitations of local search algorithms.
    
    \paragraph*{FPT-AS for Special Cases of Matroids} Let us recall some definitions. 
    A \emph{uniform matroid}  of rank $k$ is a matroid where the independent sets are the sets $S$ of cardinality at most $k$. A \emph{partition matroid}  is a matroid where we are given a partition $V_1, \dots, V_r$ of the ground set $V$ and bounds $k_1, \dots, k_r$ such that a set $S$ is independent if for all $1 \leq i \leq r$, $|S \cap V_i| \leq k_i$. 
    A \emph{laminar matroid} is given as a laminar family $V_1, \dots V_r$ of $V$, \emph{i.e.} given $V_i \neq V_j$, then either $V_i \cap V_j=\emptyset$, or $V_i \subset V_j$, or $V_j \subset V_i$, along with bounds $k_1, \dots, k_r$; a set $S \subseteq V$ is independent if for all $1 \leq i \leq r$, $|S \cap V_i| \leq k_i$. 
    Finally, a \emph{transversal matroid} is given in a family $V_1,\dots, V_k \subseteq V$, where $V_i$s are not necessarily disjoint, and a set $S=\{u_1,\cdots, u_t\}$ is independent if and only if for each element $u_i$, there exists a distinct $\phi(i)$ so that $u_i \in V_{\phi(i)}$. These simple types of matroid have been extensively studied in a large variety of contexts.
    
    A uniform matroid is a special case of a partition matroid, which is again a special case of a laminar or a transversal matroid. However, laminar matroids and transversal matroids are not inclusive of each other~\cite{Recski}. Transversal matroids were introduced in the 60s, by Edmonds and Fulkerson~\cite{EF1965} and by Mirsky and Perfect~\cite{MP1967}. They unified many results in transversal theory and are generally considered an important class of matroids, \emph{e.g.}, see~\cite{Welsh}. Laminar matroids receive much attention recently in the community of theoretical computer science, especially in the context of matroid secretary problem, \emph{e.g.}, see~\cite{BabaioffIK07,ChakrabortyL12,FSZ2015,IW2011,Soto2011}. Our FPT-AS involves these kinds of matroids.

    \begin{theorem} 
    \label{thm:mainResult}
    For every $\varepsilon>0$, we can extract an approximate kernel $V' \subseteq V$ in polynomial time so that a $(1-\varepsilon)$-approximate solution is contained in $V'$. The size of the kernel $V'$ depends on the type of the given matroid 
    $\mathcal{M}$.
    \begin{enumerate}[(i)]
        \item $|V'| \leq \frac{k}{\varepsilon}$ when $\mathcal{M}$ is a partition matroid;
        \item $|V'| \leq \frac{2k}{\varepsilon}$ when 
        $\mathcal{M}$ is a laminar matroid; 
        \item $|V'| \leq \frac{k}{\varepsilon} + k(k-1)$ when 
        $\mathcal{M}$ is a transversal matroid. 
    \end{enumerate}
    
    Furthermore, by a brute force enumeration, we can find the desired $(1-\varepsilon)$ approximation 
    in $\left(\frac{1}{\varepsilon}\right)^{O(k)}n^{O(1)}$ time for partition and laminar matroids and $\left(\frac{1}{\varepsilon}+k\right)^{O(k)}n^{O(1)}$ time for transversal matroids.   
    \end{theorem}

    Note that the result for transversal matroids has since been improved in~\cite{Kamiyama22-robust} to get a kernel of size $\frac{k}{\varepsilon}$. Besides, by a straightforward modification of our proofs in Section~\ref{sec:kernelization} (see Appendix~\ref{app:hypergraph}), we can show the following corollary. 
    \begin{corollary} Suppose that we are given a hypergraph $G=(V,E)$ with edge size bounded by a constant $\eta \geq 2$. We can compute a $(1-(\eta-1) \cdot \varepsilon)$ approximation
    using $\left(\frac{1}{\varepsilon}\right)^{O(k)}n^{O(1)}$ time for partition and laminar matroids and $\left(\frac{1}{\varepsilon}+k\right)^{O(k)}n^{O(1)}$ time for transversal matroids. 
    \label{cor:hypergraph}
    \end{corollary}
    
    Put slightly differently, when $G$ is a hypergraph with edge size at most $\eta$, we can obtain a $(1-\varepsilon)$ approximation in $\left(\frac{\eta}{\varepsilon}\right)^{O(k)}n^{O(1)}$ or $\left(\frac{\eta}{\varepsilon} + k\right)^{O(k)}n^{O(1)}$ time, depending on the type of matroid. To see the interest of this corollary, we recall that recently Manurangsi~\cite{DBLP:conf/soda/Manurangsi20} showed that if $\eta$ is unbounded, one cannot obtain an approximation ratio better than $1-1/e+\varepsilon$, assuming GAP-ETH, in FPT time (where the matroid rank $k$ is the parameter). This result holds even for the simplest uniform matroid. Thus Corollary~\ref{cor:hypergraph} implies that one can circumvent this lower bound by introducing another parameter $\eta$, even for more general matroids. 
    
    Our algorithm is inspired by that of Manurangsi~\cite{manurangsi2018note} for the case of uniform matroid. So let us briefly summarize his approach: an \emph{approximate kernel}\footnote{In the rest of the paper, we will just say kernel,  dropping the adjective. Roughly speaking, a $(1-\varepsilon)$-approximate kernel is a kernel which contains a $(1 - \varepsilon)$-approximate solution of the original problem. 
    The interest for this kind of kernel has risen recently~\cite{ FeldmannSLM20,LPRS17}.} $V'$ is first extracted from $V$, where $V'$ is simply made of the $k/\varepsilon$ vertices with the largest weighted degrees. Let $O$ be an optimal solution. Apparently, a vertex of $O$ is either part of the kernel $V'$, or its weighted degree is dominated by all vertices in $V'\backslash O$. To recover the optimal value, we can potentially use the vertices in $V'\backslash O$ to replace the vertices in $O \backslash V'$. However, there is a risk in doing this: an edge among the vertices in $V'\backslash O$ can be double-counted, if both of its endpoints are chosen to replace the vertices in $O\backslash V'$. To circumvent this issue, Manurangsi uses a random sampling argument to show that in expectation such double counting is negligible. Therefore, by the averaging principle, there exists a $(1-\varepsilon)$-approximate solution in the kernel $V'$, which can be found using brute force.
    
    To generalize the approach of Manurangsi for more general matroids, one has to answer the quintessential question: how does one guarantee that the sampled vertices, along with $O\cap V'$, are independent in $\mathcal{M}$? To overcome this difficulty, we introduce a new technique. We take the union of some number $\tau$ of matroids $\mathcal{M}$. Such a union is still a matroid, which we denote as $\tau \mathcal{M}$. We then apply a greedy algorithm on $\tau \mathcal{M}$ (based on non-increasing weighted degrees) to construct an independent set $V'$ in $\tau \mathcal{M}$. We show that such a set $V'$ is ``robust'' (see Definition~\ref{def:robustness}) in the sense that we can sample vertices from $V'$ so that they, along with $O \cap V'$, are \emph{always} independent and \emph{in expectation} cover edges of weight at least $(1-\varepsilon)$ times that of $O$. 
    
    We note that the value of $\tau$ automatically gives an upper bound on the kernel size $V'$, which is $\tau k$. Theorem~\ref{thm:greedyonUnion} shows the required scale of $\tau$,  depending on the type of the given matroid. We leave as an open question whether, for matroids more general than considered in the paper, a larger $\tau$ can always yield the kernel.
    
    In Section~\ref{sec:streaming} we consider the problem in the semi-streaming model~\cite{Mut2005}. In that context, the edges in $E$ arrive over time but we have only limited space (for instance, $O(n \cdot polylog(n)) = o(m)$) and cannot afford to store all edges in $E$. In this context we can also obtain a $(1-\varepsilon)$ approximation using $O(\frac{n k}{\varepsilon})$ space in a single pass.\footnote{Here we assume that the matroid is given in the form of an oracle, which can be accessed by the algorithm freely---this is a standard assumption in the streaming setting when matroids are involved.} The idea of using (parameterized) kernels for streaming algorithms has recently been introduced, for instance in~\cite{ChitnisCEHMMV16,McGregorTV2021}. We also show that an FPT-streaming algorithm can be derived from our ideas to get a $(1 - \varepsilon)$ approximation for a special form of maximization of a coverage function with bounded frequency (see Theorem~\ref{thm:stream-simple}, Remark~\ref{rem:coverageStreaming} and Appendix~\ref{app:stream-hypergraph} for details).
    
    \paragraph*{Local Search for General Matroids} 
    In the second part of the paper, we consider general matroids. As mentioned above, we aim for a ratio better than $1-1/e$. Combining known techniques involving LP, $3/4$ approximation can be achieved in polynomial time, as we mentioned earlier. However, the aforementioned LP-based approach involves quite heavy machinery, and is likely to be slow and hard to implement. Here we seek for simple approaches. A natural candidate is the \emph{local search} technique. However, if we try to optimize according to the original objective function $E_G(S)$ (in the literature, this is called \emph{oblivious} local search~\cite{Filmus2012a}), we show in Appendix~\ref{app:oblivious} that one cannot achieve better than $1/2$ approximation, even if we allow large exchanges. 
    
    What we do is an adaptation of a technique of Filmus and Ward~\cite{Filmus2012a}, originally designed for a general coverage function. Their idea is to optimize a special potential function, which takes the form of a linear sum of the 
    covered elements, but with the modification that elements covered multiple times have larger coefficients. 
    
    \begin{theorem} Given a general matroid $\mathcal{M}$ as constraint, we can obtain: 
    \begin{enumerate}[(i)]
        \item A $2/3-O(k\varepsilon)$ approximation algorithm using $O\left(\frac{mk}{\varepsilon}\right)$ arithmetic operations 
        and $O\left(\frac{nk}{\varepsilon}\right)$ matroid oracle calls. 
        \item A $2/3$ approximation algorithm using $O(nmk^3)$ arithmetic operations 
        and $O(n^2k^3)$ matroid oracle calls.
    \end{enumerate}
    \end{theorem}
    
    See Section~\ref{sec:local-search-2/3} for this result. 
    Here we have fast running time thanks to the fact that 
    small exchanges (of 1 element) are enough to establish the desired ratio. In Section~\ref{sec:local-search-3/4}, 
    we show that by a slight modification, we can improve the approximation ratio to $3/4$ in FPT time--precisely, in $O(k! \cdot n^{O(1)})$ time. Thus if $k=O(\log n/ \log \log n)$, we still use only polynomial time.
    
    We next consider the more general case where \emph{two} matroids 
    are imposed on the graph vertices. A solution is feasible 
    only if it is a common independent set in both matroids. 
    
    \begin{theorem}
        Given two general matroids $\mathcal{M}_1$ and $\mathcal{M}_2$ as constraints, we can obtain $\frac{2}{3}\cdot (1-\frac{1}{p+1})$ approximation, using $n^{O(p)}$ arithmetic operations and matroid oracle calls. 
        \label{thm:twoMatroids}
    \end{theorem}
    
    This result is presented in Section~\ref{local-search-2matroid}. To see the interest of Theorem~\ref{thm:twoMatroids}, observe that it implies that when the submodular function is a coverage function of frequency 2, we can obtain an approximation ratio significantly better than $1/2-\varepsilon$~\cite{Feldman2012,Lee2010a}, which is known for a general submodular function, under the constraint of two matroids. The only other special case where this ratio is improved upon that we are aware of, is when the submodular function is a weighted rank function~\cite{linhares19}, where the ratio of exactly $1/2$ is attained. 
    
    Our main tool is a characterization theorem of Lee et al.~\cite{Lee2010a} based on a certain exchange graph~\cite[Section 41.2]{Schrijver2003}, originally used for matroid intersection. The much higher running time is due to the fact that here we are obliged to use larger exchanges (of up to $\theta(p)$ elements). 
    
    \paragraph*{Some Remarks on Local Search} Considering the closeness of the approximation ratios with regard to one or two matroids, it is tempting to ask if one can do better than $2/3$ if there is just one matroid---maybe by trying larger exchanges.
    
    We show in Section~\ref{subsec:p-exch-2/3} that using the same type of potential functions~\cite{Filmus2012a} (linear sum with modified coefficients based on the number of times an edge is covered), $2/3$ approximation is the best possible even if we allow larger $O(1)$ exchanges. 
    
    Besides, another naturally related question is whether one can do better when the graphs are simpler, for instance, when the vertex degree is bounded by some constant $d$. Again we show in Section~\ref{subsec:p-exch-2/3} that the same type of potential functions cannot achieve better than $2/3+\theta(1/d)$ approximation, even if large exchanges are allowed.
    
    These examples seem to indicate a certain limitation on the power of currently known local search techniques. Other than the approach of~\cite{Filmus2012a,Filmus2014}, we are unaware of any other type of local search for coverage (or general submodular) functions. To do better than $2/3$ (say up to $3/4$), a significantly different new idea may be required.

\section{Kernelization Framework}
    \label{sec:kernelization}
    
    In this section, we give a general framework to construct the kernel by a greedy procedure and show how such a kernel contains a $(1-\varepsilon)$-approximate solution.

    \begin{definition}
    \label{def:robustness}
    Let $\mathcal{M}= (V, \mathcal{I})$ be a matroid with weights $\omega: V \rightarrow \mathbb{R}^+$. 
    We say $V' \subseteq V$ is $t$-robust if given any base $O \in \mathcal{I}$, there is a bijection from the elements $u_1,\cdots,u_r \in O \backslash V'$ to subsets $U_{u_1},\cdots,U_{u_r} \subseteq V' \backslash O$ so that 
    \begin{enumerate}[(i)]
        \item the $U_{u_i}$s are mutually disjoint and $|U_{u_i}|=t$, 
        \item all elements in $U_{u_i}$ have weights 
        no less than $u_i$, 
        \item by taking an arbitrary element $u'_i \in U_{u_i}$ for all $i$, $(V' \cap O) \cup \{u'_i\}_{i=1}^{r}$ is a base in $\mathcal{M}$. 
    \end{enumerate}
    
    \end{definition}

    \begin{figure}
        \centering
        \begin{tikzpicture}
            \node (O) at (-1,0) {$O$};
            \node [rectangle, draw] (Oin) at (-0.3,-2) {$O^{in}$};
            \node (Oout) at (1,0) {$O^{out} :$};
            
            \draw [->] (O) -- (Oin);
            \draw [->] (O) -- (Oout);

            \node (u1) at (2,0) {$u_1$};
            \node (up) at (3,0) {$\cdots$};
            \node (ur) at (4,0) {$u_r$};
            \draw [draw=black] (0.4,-0.3) rectangle (4.3,0.3);

            \draw [draw=black] (-1,-0.5) rectangle (5.2,-4);
            \node (Vp) at (0,-1) {$V':$};
            
            \node (Uu1) at (1.5, -1) {$U_{u_1}:$};
            \node (v1) at (1.5, -1.5) {$v_1$};
            \node (v2) at (1.5, -2) {$v_2$};
            \node (v3) at (1.5, -2.5) {$v_3$};
            \draw [draw=black] (1,-0.7) rectangle (2,-2.8); 
            \draw [->] (u1) -- (Uu1);
            
            \node at (3, -1.75) {$\cdots$};    
            
            \node (Uur) at (4.5, -1) {$U_{u_r}:$};
            \node (v7) at (4.5, -1.5) {$v_7$};
            \node (v8) at (4.5, -2) {$v_8$};
            \node (v9) at (4.5, -2.5) {$v_9$};
            \draw [draw=black] (4,-0.7) rectangle (5,-2.8); 
            \draw [->] (ur) -- (Uur);     
            
            \node at (0, -3.5) {$v_{10}$};
            \node at (1, -3.2) {$v_{11}$};
            \node at (2, -3.7) {$v_{12}$}; 
            \node at (3, -3.5) {$\cdots$}; 
        \end{tikzpicture}
        \caption{Representation a robust decomposition, for $t = 3$, with $O^{in} = V' \cap O$ and $O^{out} = O\backslash O^{in}$. We have that (i) the $U_{u_i}$s are disjoint, (ii) $\forall\, 1 \leq i \leq r, \forall\, v \in U_{u_i}, \omega(u_i) \leq \omega(v)$, and (iii) taking an arbitrary element $u'_i \in U_{u_i}$ for all $i$, $(V' \cap O) \cup \{u'_i\}_{i=1}^{r}$ is independent in $\mathcal{M}$.}
    \end{figure}
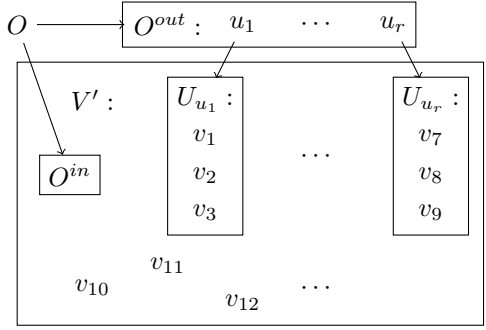

    We next recall the definition of matroid union. 
    
    \begin{definition} Suppose that $\mathcal{M}=(V, \mathcal{I})$ is a matroid. Then 
    we can define $\tau \mathcal{M} =(V, \mathcal{I}_{\tau})$ 
    as the union of $\tau$ matroids $\mathcal{M}$, as follows: $S \in \mathcal{I}_{\tau}$ if $S$ can 
    be partitioned into $S_1\cup \cdots \cup S_{\tau}$ so that each $S_i \in \mathcal{I}$. 
    \end{definition}
    
    Recall that the union of matroids is still 
    a matroid and here the rank of $\tau \mathcal{M}$ is at most $\tau$ times the rank of $\mathcal{M}$, \emph{e.g.}, see~\cite[Chapter 42]{Schrijver2003}. 
    We can now state our main theorem. 
    
    \begin{theorem}
    \label{thm:greedyonUnion}
    Let $\mathcal{M}=(V, \mathcal{I})$ be a 
    matroid with weights $\omega:V \rightarrow \mathbb{R}^+$ and rank $k$. Consider the following greedy procedure on $\tau \mathcal{M}=(V,\mathcal{I}_{\tau}) $ to construct $V'$: initially $V'=\emptyset$. Process the elements in $V$ by non-increasing weights $\omega$. For each element $u$, if $V'+u \in \mathcal{I}_{\tau}$, add $u$ into $V'$, otherwise, ignore it. The final $V'$ is $t$-robust 
    \begin{enumerate}[(i)]
        \item if $\mathcal{M}$ is a partition matroid and $\tau \geq t$,
        \item if $\mathcal{M}$ is a laminar matroid and $\tau \geq 2t$, 
        \item if $\mathcal{M}$ is a transversal matroid and $\tau \geq t+k-1$. 
    \end{enumerate}
    \end{theorem}
    
    Notice that the rank of the matroid $\tau \mathcal{M}$ gives an upper-bound on the size of $V'$. The next section will give the proof of this theorem for each type of matroid considered.  
    In the following we show how it can be used to construct the $(1-\varepsilon)$ approximation.  
    
    Let the weight $\omega:V \rightarrow \mathbb{R}^+$ be the weighted degrees in the graph $G=(V,E)$, that is, $\omega(u) = \deg_w(u)$. Apply Theorem~\ref{thm:greedyonUnion} by setting $t =\frac{1}{\varepsilon}$. Then $V'$ is $\frac{1}{\varepsilon}$-robust. Note that we suppose that $\frac{1}{\varepsilon}$ is an integer, otherwise we could take $t =\lceil\frac{1}{\varepsilon}\rceil$.
    
	Based on $V'$, we create a new graph 
	$G' = (V', E')$, where an original edge $e=\{u,v\}$ is retained in $E'$ if 
	both of its endpoints are in $V'$. In case only one endpoint, say $u$ is in $V'$, we add a self-loop to $u$ in $E'$ to represent this edge.

	\begin{lemma} 
	    Suppose that $V'$ is the constructed set that is $\frac{1}{\varepsilon}$-robust. Then $V'$ contains a set $S$ such that $S \in \mathcal{I}$ and $E_G(S) \geq (1 - \varepsilon) \cdot E_G(O)$ where $O$ denotes an optimal solution of the problem.
	\end{lemma}

	\begin{proof}
	    Let $O \in \mathcal{I}$ be an optimal solution. We denote $O^{in} = O \cap V'$, $O^{out} = O \backslash O^{in}$. Then by $\frac{1}{\varepsilon}$-robustness, we have mutually disjoint sets $U_v \subseteq V' \backslash O$ for each $v \in O^{out}$, each of size $\frac{1}{\varepsilon}$. We set $U= \cup_{v \in O^{out}}U_v$. We construct a set $S \subseteq V'$ as follows: $S$ is initialized as $O^{in}$. Then from each set $U_v$, for all $v \in O^{out}$, pick an element at random and add it into $S$. By definition of $\frac{1}{\varepsilon}$-robustness, $S$ is independent in $\mathcal{M}$. 
	    
	    Next we will show that \[\mathbb{E}[E_G(S)] \geq (1 - \varepsilon) \cdot \mathbb{E}[E_G(O)].\]
	    Let $U^* = S \backslash O^{in}$, i.e. 
	    those elements that are added into $S$ 
	    randomly. 
	    First, we have that:
	    \[E_G(S) = E_G(O^{in}) + E_G(U^*) - E_G(O^{in}, U^*).\]
	    We bound $\mathbb{E}[E_G(O^{in}, U^*)]$ as follows. By construction, $\mathbb{P}[u \in U^*] = \varepsilon$ for all $u \in U$. Then, 
	    \[\mathbb{E}[E_G(O^{in}, U^*)] = \sum_{u \in U}\sum_{v \in O^{in}} w(\{u, v\}) \cdot \mathbb{P}[u \in U^*] = \varepsilon \sum_{u \in U}\sum_{v \in O^{in}} w(\{u, v\}) \leq \varepsilon \cdot E_G(O^{in}).\]
	    Furthermore, the value $\mathbb{E}[E_G(U^*)]$ can be rearranged as follows: 
	    \begin{align*}
	        \mathbb{E}[E_G(U^*)] &= \mathbb{E}\left[\sum_{u \in U^*}\left(\deg_w(u) - \frac{1}{2} \sum_{v \in U^*\backslash \{u\}}w(\{u,v\})\right)\right]\\
	        =&\, \sum_{u \in U}\left(\deg_w(u) \cdot \mathbb{P}[u \in U^*] - \frac{1}{2} \sum_{v \in U \backslash \{u\}}  w(\{u, v\}) \cdot \mathbb{P}[u \in U^* \wedge v \in U^*]\right)\\
	        \geq&\, \sum_{u \in U}\left(\deg_w(u) \cdot \varepsilon - \frac{1}{2} \sum_{v \in U \backslash \{u\}}  w(\{u, v\}) \cdot \varepsilon^2\right) \geq \varepsilon \cdot (1 - \varepsilon/2) \left(\sum_{u \in U} \deg_w(u)\right),
	    \end{align*}
	    where the first inequality comes from the fact that $\mathbb{P}[u \in U^* \wedge v \in U^*] \leq \mathbb{P}[u \in U^*] \cdot \mathbb{P}[v \in U^*]$.
	    
	    Recall that by robustness, for all $u \in O^{out}$, the elements of $U_{u}$ have weighted degrees no less than that of $u$. Therefore, 
	    \begin{align*}
	        \mathbb{E}[E_G(U^*)] \geq \varepsilon \cdot (1 - \varepsilon/2) \left(\sum_{u \in O^{out}}\sum_{v \in U_{u}}\deg_w(v)\right) &\geq \varepsilon \cdot (1 - \varepsilon/2) \left(\sum_{u \in O^{out}}\frac{1}{\varepsilon} \cdot \deg_w(u)\right)\\
	        &\geq (1 - \varepsilon/2) \cdot E_G(O^{out}).
	    \end{align*}

	    As a result, we get:
	    \[\mathbb{E}[E_G(S)] \geq E_G(O^{in}) + (1 - \varepsilon/2) \cdot E_G(O^{out}) - \varepsilon \cdot E_G(O^{in}) \geq (1 - \varepsilon) \cdot E_G(O).\]
	    By averaging principle, there exists $S \subseteq V'$ such that $S \in \mathcal{I}$ and $E_G(S) \geq (1 - \varepsilon) \cdot E_G(O)$.
	\end{proof}
	
\section{Proof of Theorem~\ref{thm:greedyonUnion}}
    \label{sec:ker_meth}
	\subsection{Partition Matroids}
        \label{subsec:partition}
        
        
        Consider a partition matroid $\mathcal{M} = (V, \mathcal{I})$ defined by a partition $V_1, \dots, V_r$ of $V$ and bounds $k_1, \dots, k_r$. Given $t \in \mathbb{N}$, we take in each set $V_i$ of the partition the $\min\{|V_i|, t \cdot k_i\}$ elements having the largest weighted degrees. We denote these extracted sets as $V_i'$ and their union as $V'$. Clearly, this is the same as the greedy algorithm stated in Theorem~\ref{thm:greedyonUnion} applied on $t\mathcal{M}$.
        
        To see that $V'$ is $t$-robust, let $O \in \mathcal{I}$ be a base. We denote $O^{in} = O \cap V'$, $O^{out} = O \backslash O^{in}$, $O_i = O \cap V_i$, $O_i^{in} = O \cap V'_i$, $O_i^{out} = O_i \backslash O_i^{in}$, and we set $\overline{U}_i \subseteq V'_i \backslash O^{in}_i$ as an arbitrary subset of cardinality $t \cdot |O^{out}_i|$, for $1 \leq i \leq r$ (it is possible as $O_i \neq O_i^{in}$ implies that $|V'_i| = t \cdot k_i$). We then partition $\overline{U}_i$ into $U^{1}_i$, $\dots$, $U^{|O^{out}_i|}_i$, each one of size $t$. It is easy to verify that the generated sets $\{U^{j}_{i}\}_{i,j}$ satisfy the three conditions stated in Definition~\ref{def:robustness}. 
        
	
    \subsection{Laminar Matroids}
        \label{subsec:laminar}
        Recall that a laminar matroid $\mathcal{M}=(V, \mathcal{I})$ is given as a laminar family $V_1, \dots, V_r$ along with bounds $k_1, \dots, k_{r}$. Without loss of generality, we can assume that $V=V_0$ with bound $k_0 = k$ (being the rank of $\mathcal{M}$) is a member in the family. Furthermore, we can also assume that each vertex $v \in V$ by itself is also a member $V_i=\{v\}$ with bound $k_i=1$ in this family. 
        
        Such a laminar matroid $\mathcal{M}$ can be naturally associated with a laminar tree $T$ where each tree node $T_i = (V_i, k_i)$ corresponds to $V_i$, and the structure of the tree reflects the inclusion relationship of the members $V_i$ in the laminar family. In such a tree $T_0 = (V,k)$ corresponds to the root of the tree. 
        
        For ease of our study, we will assume that such a tree $T$ is 
        binary (so in total $T$ contains $2n-1$ nodes, with $n = |V|$). Such an assumption can be easily justified by adding more sets $V_i$ into the laminar family with the appropriately defined bounds $k_i$. 
        
    	In the following, the elements of $\{T_i = (V_i, k_i)\}_{0 \leq i \leq 2n-2}$ will be referred as ``nodes'', whereas the elements of $V$ will be referred as ``vertices''.  We are given $t\in \mathbb{N}$. To choose the vertices that are to be added into the kernel, we employ the following greedy procedure. We process the vertices in non-increasing order with respect to their weighted degrees. At the beginning, $V'$ is empty. When we consider a new vertex $v$, if for all $i$ such that $v \in V_i$, we have $|V' \cap V_i| < 2t \cdot k_i$, then $v$ is added to $V'$, otherwise $v$ is simply ignored. This procedure is equivalent to the greedy algorithm described in Theorem~\ref{thm:greedyonUnion} applied on $2t\mathcal{M}$.
    	
        A node $T_j$ of the tree $\{T_i = (V_i, k_i)\}_{0 \leq i \leq 2n-2}$ is called \emph{saturated} if $|V' \cap V_j| = 2t \cdot k_j$. Let $O$ be an arbitrary solution. As in the previous subsection we will use the notations $V'_i = V_i \cap V'$, $O^{in} = V' \cap O$, and $O^{out} = O \backslash V'$. In the following, we say that a vertex or a set of vertices is ``contained'' in a tree node $T_i$ if they are part of $V'_i$ (equivalently, the leaves corresponding to these elements of $V'$ are in the subtree of root $T_i$).
    	
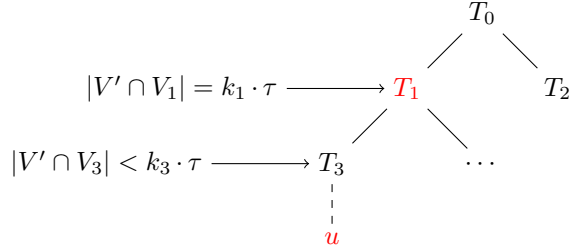
\begin{figure}
        	\centering
            \begin{tikzpicture}
                \node (V1) at (0,0) {$T_0$};
                \node [red] (V2) at (-1,-1) {$T_1$};
                \node (V3) at (1,-1) {$T_2$};
                
                \draw [-] (V1) -- (V2);
                \draw [-] (V1) -- (V3);
                
                \node (V4) at (-2,-2) {$T_3$};
                \draw [-] (V2) -- (V4);
                
                \node (V5) at (0,-2) {$\cdots$};
                \draw [-] (V5) -- (V2);
                
                \node [red] (u) at (-2,-3) {$u$};
                \draw [-,dashed] (u) -- (V4);
                
                \node (C) at  (-4, -1) {$|V' \cap V_1| = k_1 \cdot \tau$};
                \draw [->] (C) -- (V2);

                \node (D) at  (-5, -2) {$|V' \cap V_3| < k_3 \cdot \tau$};
                \draw [->] (D) -- (V4);
            \end{tikzpicture}
            \caption{\label{fig:blocking-node}The node $T_1$ is the deepest in the tree blocking $u$ (preventing it to be in the kernel).}
         \end{figure}

    	For every element $v \in V$, there exists a leaf $T_{i_v}$ in the laminar tree such that $V_{i_v} = \{v\}$, and we have a unique path from the root $T_0$ to $T_{i_v}$. If a vertex $v \in O$ is not in $V'$, it means that some node along the path from $T_0$ to $T_{i_v}$ was already saturated when $v$ was processed: the \emph{blocking node} of $v$ is the deepest saturated node along this path (see Figure~\ref{fig:blocking-node}). For each node $T_i$, we denote by $B_i$ the set of vertices of $O$ that are blocked by the node $T_i$, and we set $b_i = |B_i|$. Then, $b_i = 0$ when $T_i$ is not saturated. Moreover, the $B_i$s are mutually disjoint and $\bigcup B_i = O^{out}$.

    	Then for each vertex $v\in O^{out}$, we construct a set $\overline{U}_v$ of at least $t$ vertices drawn from $V'$. Then an arbitrary subset $U_v \subseteq \overline{U}_v$ of $t$ vertices is retained. We will argue that the generated sets $U_v$s ensure the robustness.
    	    
    	    \paragraph*{Constructing the sets $\overline{U}_v$s for all $v \in O^{out}$}
    	    We want the constructed sets $\overline{U}_v$s to satisfy the following three properties. 
    	    
    	    \begin{enumerate}
    	       
    	    \item[(i)] The sets $\overline{U}_v$s are mutually disjoint and are drawn from $V'\backslash O^{in}$. 
    	    \item[(ii)] For each $v \in O^{out}$ and each $u \in \overline{U}_v$, 
    	    $\deg_w(u) \geq \deg_w(v)$. 
    	    
    	    \item[(iii)] Choosing an arbitrary $\overline{v} \in \overline{U}_v$ for each $v \in O^{out}$, the set $S = O^{in} \cup \{\overline{v}\}_{v \in O^{out}}$ 
    	    is independent in the laminar matroid $\mathcal{M}$. 
    	    \end{enumerate}
    	    
    	    The formal algorithm for constructing the sets $\{\overline{U}_v\}_{v \in O^{out}}$ is given in Algorithm~\ref{construct-S}. Here we give the 
    	    intuition behind it. 
    	    
    	    To guarantee Property (i), we first mark all elements in $O^{in}$ as unusable. Then, each $\overline{U}_v$ is chosen among the usable 
    	    vertices. Once a set $\overline{U}_v$ is allocated, all its vertices will be marked as unusable.
    	    
    	    To guarantee Property (ii), first recall that each vertex $v \in O^{out}$ has a corresponding blocking node $T_i$ (and $v \in B_i$). By our greedy procedure to build the kernel $V'$, we know that there exist $2t \cdot k_i$ vertices $u$ in the set $V'_i$, all of whom are contained in $T_i$ and $\deg_w(u) \geq \deg_w(v)$. What we do is to choose a deepest blocking node $T_i$ and to process one of its vertex $v \in B_i$ (Lines~6-7). 
    	    As we will show later (Claim~\ref{cla:alwaysFind}), such a blocking node must contain at least $2t$ usable vertices. We climb down the tree from the blocking node $T_i$ until we reach a node $T_j$ neither of whose child nodes contains more than $t$ usable vertices (Lines~9-10). 
    	    Recall that our tree is binary, as a result, the number of usable vertices 
    	    contained in $T_j$ is between $t$ and $2t-2$. All these usable vertices 
    	    constitute a new set $\overline{U}_v$ and then are marked as unusable. 
    	    
    	    How to guarantee Property (iii) is the most tricky part of the algorithm. Recall that we will choose an arbitrary vertex from $\overline{U}_v$ to construct a solution $S$ stated in (iii). Apparently we have no control over the choice of the arbitrary vertex from $\overline{U}_v$, nonetheless, we need to ensure that $S$ does not violate any of the rank constraints $k_i$. What we do is to associate a variable $s_i$ with each tree node $T_i$. This variable indicates how many vertices contained in $T_i$ will \emph{certainly} be part of $S$, according to $O^{in}$ and the sets $\overline{U}_v$s that have been constructed so far. Once $s_i$ is set to $k_i$, it is a warning that we should not use any more remaining usable vertices contained in $T_i$ to construct the future sets $\overline{U}_v$. 
    	    
    	    Initially, $s_i = |V'_i \cap O^{in}|$. Each time that we have decided on a tree node $T_j$ to form a new set $U_v$ (Lines 9-12), we increase the value of $s_j$ from $T_j$ all the way up to the root (Lines~13-14). If any node $T_j$ has its variable $s_j=k_j$, we say such a node is fully-booked and we mark all its (remaining) usable vertices as unusable (Lines~15-16). 
    	    
    	    \begin{algorithm}
        	\caption{Algorithm constructing the sets $\overline{U}_v$} \label{construct-S}
        	\begin{algorithmic}[1]
        	\State $\forall v \in O \backslash V',\, \overline{U}_v \gets \emptyset$
        	\State $\forall 1 \leq i \leq 2n-2,\, s_i \gets |V'_i \cap O^{in}|$
        	\State the elements of $V' \backslash O^{in}$ are marked as usable
        	\State the elements of $O^{in}$ are marked as unusable
        	\While{there exists a set $B_i$ which is not empty}
        	    \State let $T_i$ be one of the deepest nodes such that $B_i \neq \emptyset$
        	    \State let $v \in B_i$ be an arbitrary vertex
        	    \State $B_i \gets B_i - v$
        	    \While{$T_i$ has a child $T_j$ containing at least $t$ usable vertices in $V'_j$}
        	        \State $i \gets j$
        	    \EndWhile
        	    \State set $\overline{U}_v$ as the set of usable elements in $V'_i$
        	    \State mark all the elements of $\overline{U}_v$ as unusable
        	    \For{all nodes $T_j$ on the path from $T_i$ to the root of the tree}
        	        \State $s_j \gets s_j + 1$
        	        \If{$s_j = k_j$} \Comment in that case, we say that $T_j$  is \emph{fully-booked}
        	            \State mark all the elements of $V'_j$ as unusable
        	        \EndIf
        	    \EndFor
        	\EndWhile
        	\end{algorithmic}
        	\end{algorithm}
        	
        	We want to show that Algorithm~\ref{construct-S} manages to build the sets $\overline{U}_v$s of size at least $t$ satisfying the aforementioned properties. 
        	
        	\begin{claim} \label{claim:usable}
        	    At any time during the execution of Algorithm~\ref{construct-S}, for a saturated node $T_i$ that is not a descendant of a fully-booked node (no node above it is fully-booked), the number of usable vertices in $V'_i$ is at least $2t \cdot (k_i - s_i)$.
        	\end{claim}
        	
        	\begin{proof}
        	    As $T_i$ is saturated, $V'_i$ contains exactly $2t \cdot k_i$ vertices. The unusable vertices in $V'_i$ fall into three categories:
        	    \begin{enumerate}[(i)]
        	        \item the vertices in $V'_i \cap O^{in}$ (see Line~4), but that are not contained in any fully booked descendent node of $T_i$,
        	        \item the vertices made unusable during the allocation of a set $\overline{U}_v$ (see Line~12), but that are not contained in any fully booked descendent node of $T_i$,
        	        \item the vertices that are contained in a fully-booked descendent node of $T_i$ (see Line~16).
        	    \end{enumerate}
        	    The vertices $v_1,\dots,v_{l_1}$ in the first category each has a contribution of $+1$ in the value of $s_i$. Then, we can observe that an allocated set $\overline{U}_v$ is either entirely contained in a fully-booked node or has no element at all in any fully-booked node. Let us denote $\overline{U}_{v_1}, \dots, \overline{U}_{v_{l_2}}$ the allocated sets contained in $T_i$ that are not contained in any fully-booked node. In addition, by construction, each set $\overline{U}_{v_j}$ contains at most $2t - 2$ vertices (otherwise we would be able to go deeper in the binary tree for the allocation, see Lines~9-10, because at least one child would contain at least $t$ usable elements). Each set $\overline{U}_{v_j}$ has a contribution of $+1$ in the value of $s_i$. Finally, among the fully-booked nodes in the subtree of root $T_i$, we consider the nodes $T_{i_1},\dots,T_{i_{l_3}}$ that are inclusion-wise maximal (\emph{i.e.} the roots of the fully-booked parts of the subtree). A fully-booked subtree of root $T_j$ has to bring a $+k_j$ contribution to $s_i$ (otherwise it would not be fully-booked, and observe that a set $\overline{U}_v$ is included in at most one such fully-booked maximal node), and is making at most $2t \cdot k_j$ vertices of $V'_i$ unusable (the worst case being that $T_j$ was also a saturated node). We also have the equality $s_i =  l_1 + l_2 + \sum_{j = 1}^{l_3} k_{i_j}$. As a result, we have at most $l_1 + (2t - 2) \cdot l_2 + \sum_{j = 1}^{l_3} 2t \cdot k_{i_j} \leq 2t \cdot s_i$ unusable vertices in $V'_i$.
        	\end{proof}
        	
        	\begin{claim} \label{claim:ineg_k}
        	    At any time during the execution of the algorithm, for all $i \in \llbracket 1, 2n-2 \rrbracket $, we have $k_i \geq s_i + b_i + \sum_{\text{$T_j$ below $T_i$}} b_j$ as an invariant.
        	\end{claim}
        	
        	\begin{proof}
        	    As $O \in \mathcal{I}$, these inequalities hold at the beginning of Algorithm~\ref{construct-S}. To see this, note that $s_i = |V_i \cap O^{in}|$ and $b_i + \sum_{\text{$T_j$ below $T_i$}}b_j \leq |V_i \cap O^{out}|$. For the induction step, observe that Line~6 guarantees that the node $T_i$ selected is the only one with a non-zero $b_i$ value in the subtree of root $T_i$. As a result, when the set $\overline{U}_v$ is allocated, for the nodes $T_j$ between $T_i$ and the allocated node, the augmentation by one of the values $s$ is not an issue because these nodes are chosen to be non-fully-booked, \emph{i.e.} $k_j > s_j$. For each node $T_j$ above $T_i$, the value $s_j$ is increased by one but as $b_i$ was decreased by one, the total value $s_j + b_j + \sum_{\text{$T_{j'}$ below $T_j$}} b_{j'}$ remains unchanged.
        	\end{proof}
        	
        	\begin{claim} \label{cla:alwaysFind}
        	    If $B_i \neq \emptyset$, $T_i$ contains at least $2t$ usable 
        	    vertices. Consequently, Algorithm~\ref{construct-S} (Lines 6-11) always builds the set $\overline{U}_v$ of size at least $t$.
        	\end{claim}
        	
        	\begin{proof}
        	    In fact, as $B_i \neq \emptyset$, $b_i$ is still non-zero, and by Claim~\ref{claim:ineg_k} we get $k_i - s_i \geq b_i$, so $V'_i$ contains at least $2t$ usable vertices because of Claim~\ref{claim:usable}. Then the set $\overline{U}_v$ can be built as required, containing at least $t$ elements.
        	\end{proof}
        	
        	The above claim lower-bounds the size of each $\overline{U}_v$. The fact that the constructed $\overline{U}_v$s are mutually disjoint follows from the algorithm (Line~12). 
        	Now we want to show that the $\overline{U}_v$s have the desired properties regarding independence and weighted degrees.
        	
        	\begin{claim} \label{claim:indep}
    	        At any time during the execution of Algorithm~\ref{construct-S}, if we build a set $S$ by taking the elements of $O^{in}$ and one arbitrary element in each set $\overline{U}_v$ that has already been constructed, then $S$ is independent. Moreover, for a node $T_i$ that does not contain only unusable vertices in $V'_i$, any arbitrary choice of elements in the $\overline{U}_v$s will lead to the equality $|S \cap V'_i| = s_i$.
        	\end{claim}
        	
        	\begin{proof}
        	    We proceed by induction. These properties are clearly satisfied at the beginning of the algorithm (because then $S = O^{in}$). Now suppose that these properties hold at some time, and then we allocate a new set $\overline{U}_{v'}$ for some $v' \in O^{out}$. Let $S$ be made of $O^{in}$ and an arbitrary choice for the $\overline{U}_v$s that were constructed so far (excluding $\overline{U}_{v'}$). By the induction hypothesis, $S \in \mathcal{I}$. The vertices of $\overline{U}_{v'}$ are supposed to be usable, so the nodes containing them are not fully-booked and these nodes contain usable vertices. By induction on the second part of the claim, a usable element in such a node $T_j$ can be selected, as any choice for the other $\overline{U}_v$s will use exactly $s_j < k_j$ vertices of the laminar constraint of that node. Therefore any vertex $u \in V_{v'}$ added to $S$ does not cause any constraint to be violated, and $S \cup \{u\} \in \mathcal{I}$. Let $T_i$ be the node used for the allocation at Line~11 of Algorithm~\ref{construct-S}. All the nodes in the subtree of root $T_i$ will be subsequently ignored by the algorithm, as all the nodes inside it are marked as unusable. The values $s_j$ of the nodes $T_j$ in that subtree are not updated by the algorithm, but it is not an issue given that the second part of the claim does not affect them. The nodes above $T_i$ are updated, and it is true that for any vertex chosen in $\overline{U}_{v'}$, that vertex will count in the laminar inequalities for these nodes as a $+1$. This concludes the induction.
        	\end{proof}
        	
        	\begin{claim} \label{claim:degw}
        	    For all $v \in O \backslash V'$, for all $u \in \overline{U}_v$, it holds that $\deg_w(u) \geq \deg_w(v)$.
        	\end{claim}
        	
        	\begin{proof}
        	    By construction, $\overline{U}_v \subseteq V'_{i}$ where $T_{i}$ is the blocking node of $v$. As $T_{i}$ is the blocking node of $v$, all the elements in $V'_{i}$ have a larger weighted degree than $v$.
        	\end{proof}
        	
        	Finally we construct the sets $U_v$s by choosing arbitrarily $t$ vertices from $\overline{U}_v$. By Claims~\ref{cla:alwaysFind},~\ref{claim:indep}, and~\ref{claim:degw}, they satisfy the properties of robustness in Definition~\ref{def:robustness}.

    \subsection{Transversal Matroids}
        \label{subsec:transversal}
        Recall that a transversal matroid $\mathcal{M} = (V, \mathcal{I})$ can be represented as a bipartite graph $G=(A \cup V, E)$ with $A=\{A_1,\cdots, A_k\}$ representing the transversal sets. A subset $V' \subseteq V$ is independent in $\mathcal{M}$ if and only if there is a matching where all of $V'$ are matched to some subset of $A$. Let $t \in \mathbb{N}$. The matroid union $(t+k-1)\mathcal{M}$ can be regarded as making the capacity of each vertex $A_i$ in $A$ increased to $t+k-1$ (equivalently, create $t+k-1$ copies of each $A_i$ and modify the edge set $E$ accordingly). Our algorithm is as follows. Again process the vertices in non-increasing order of their weighted degrees. We start with an empty matching, and we maintain a matching throughout the execution of the algorithm. For each new vertex $v \in V$, try to find an augmenting path so that it can be matched. If we cannot find such a path, $v$ is discarded. At any time during the execution of the algorithm, the current kernel $V' \subseteq V$ is simply the set of vertices in $V$ that are matched in the current matching. We can observe that a vertex in $V'$ cannot be evicted once it belongs to $V'$. In the following, we write $V'_i \subseteq V'$ to denote the vertices in $V$ that are matched to $A_i$ in the current kernel $V'$. 
        
        First we argue that our procedure is the same as the greedy described in Theorem~\ref{thm:greedyonUnion} applied on $(t+k-1)\mathcal{M}$. We need to show that a vertex $v$, if discarded, is spanned by vertices in $V'$ that arrived earlier than it. To see this, observe that, at the moment $v$ arrives, $V'$ is independent in $(t+k-1)\mathcal{M}$. Moreover, as we cannot find an augmenting path when $v$ is added to $V'$, is means that $V' + v$ is not independent, \emph{i.e.} $v$ is spanned by $V'$ and this holds until the end of the algorithm. 
        
        Now let us consider the robustness. Let $O =\{o_1, \dots, o_k\}$ be an arbitrary base in $\mathcal{M}$. We can assume that $o_i$ is assigned to $A_i$ for all $i$ in the corresponding matching. For an element $o_i \in O \backslash V'$, when it was discarded, some $V'_i \subseteq V'$ elements (exactly $t + k -1$) that arrived earlier were already assigned to $A_i$. As no augmenting path could go through $A_i$ when $o_i$ was discarded, the set of elements assigned to $A_i$ would not have changed till the end: otherwise, that would mean that at some point an augmenting path passed through $A_i$, which is not possible as $o_i$ was discarded because no augmenting path passing through $A_i$ was found at that point. As a result the $t + k - 1$ elements of $V'_i$ assigned to $A_i$ are all of weighted degrees larger than that of $o_i$. As $|V'_i \cap O| \leq |V' \cap O| < k$ there remain at least $t$ elements of $V'_i$ that can be used to build a set $U_{o_i}$ of cardinality $t$ as stated in Definition~\ref{def:robustness}.
        
\section{Streaming Algorithms}
    \label{sec:streaming}
    In this section, we turn our algorithms into streaming form. 
    
    First, we show that it is easy to compute a $(1-\varepsilon)$ approximation in two passes, using $O(n+\tau^2)$ space ($\tau$ depends on the type of matroids involved, as defined in Theorem~\ref{thm:greedyonUnion}). In the first pass, we compute the weighted degrees $\deg_w(v)$ 
    of all vertices $v$ to define the kernel $V' \subseteq V$. This requires $O(n)$ space. In the second pass, we retain a subset of edges $E' \subseteq E$, those both of whose end-points are in $V'$. Clearly, $|E'|= O(\tau^2)$. 
    Using $E'$, we can compute the exact value $E_{G}(S) = \sum_{v \in S}\deg_w(v) - \sum_{e \in (S \times S) \cap E'}w(e)$ for each feasible independent set $S \subseteq V'$. Then an enumeration of all such sets gives the desired $(1-\varepsilon)$ approximation. 
    
    We now explain how to achieve the same goal in one pass, at the expense of higher space requirement. 

    \begin{theorem}
        In the edge arrival streaming model (each edge appearing exactly once in the stream), one can extract a $(1-\varepsilon)$-approximate solution of the matroid-constrained maximum vertex cover using $O(\frac{nk}{\varepsilon})$  variables for uniform, partition, laminar, and transversal matroids.
    \end{theorem}
    
    \begin{proof}
        Let $\varepsilon > 0$. 
        During the streaming phase, we keep track of the weighted degrees of all the vertices, as well as for each vertex $v$ the set of the $\frac{2k}{\varepsilon}$ edges incident to $v$ that have the largest weight. We denote the set of memorized edges as $E'$.
        
        Then, we can choose, depending on the type of matroid, the value $\tau$ corresponding to the right type of matroid (as prescribed in Theorem~\ref{thm:greedyonUnion}) for the parameter $\frac{\varepsilon}{2}$ and we build the kernel $V'$ that is supposed to contain a $(1 - \frac{\varepsilon}{2})$ approximation of the maximum cover. However, we do not know all the edges between the elements in $V'$, as only the $\frac{2k}{\varepsilon}$ heaviest incident edges are known for each vertex. 
        
        We will compute the value of $S$ \emph{pretending} that the edges 
        in $((S \times S)\cap E)\backslash E'$ are not present. Precisely, for each set $S\subseteq V$, we define
        
        $$\tilde{E}_G(S) = \sum_{v \in S} \deg_w(v) - \sum_{e \in (S\times S)\cap E'}w(e) = E_G(S) + \sum_{e \in ((S\times S)\cap E) \backslash E'}w(e).$$
        
        Notice that $\tilde{E}_G(S) \geq E_G(S)$. Let $S^* \subseteq V'$ be the independent set reaching the maximum $\tilde{E}_G(S^*)$. This set $S^*$ will be our final output. We next lower-bound its real value $E_G(S^*)$. 
        
        Let $O$ denote the original optimal solution (with respect to the entire graph), and $S'$ denote the optimal vertex cover in the kernel $V'$ (also with respect to the entire graph), so that $S' \subseteq V'$, $S' \in \mathcal{I}$, and $E_G(S') \geq (1 - \frac{\varepsilon}{2}) \cdot E_G(O)$. Then 
        $$\tilde{E}_G(S^*) \geq \tilde{E}_G(S') \geq E_G(S') \geq \left(1-\frac{\varepsilon}{2}\right) \cdot E_G(O).$$
        
        To compare the real value of $E_G(S^*)$ with $\tilde{E}_G(S^*)$, we just need to compute the total weight of the edges in $((S^* \times S^*)\cap E)\backslash E'$: 
        
        \begin{align*}
            \sum_{(u,v) \in ((S^* \times S^*)\cap E)\backslash E'} w(u,v) &= \frac{1}{2}\sum_{v \in S^*}\left( \sum_{u \in S^* : (u, v) \in E \backslash E'} w(u,v)\right)\\
            &\leq \frac{1}{2}\sum_{v \in S^*} k \cdot \deg_w(v) \cdot \frac{\varepsilon}{2k}\\ 
            &= \frac{\varepsilon}{4} \sum_{v \in S^*} \deg_w(v) \leq \frac{\varepsilon}{2} \cdot \tilde{E}_G(S^*),
        \end{align*}
        where the first inequality comes from the fact that the edges that are not among the $\frac{2k}{\varepsilon}$ heaviest edges incident on $v$ must be of weight at most $\deg_w(v) \cdot \frac{\varepsilon}{2k}$. Therefore the real value $E_G(S^*)$ is at least $(1 - \frac{\varepsilon}{2}) \cdot \tilde{E}_G(S^*) \geq (1 - \frac{\varepsilon}{2})^2 \cdot E_G(O) \geq (1 - \varepsilon) \cdot E_G(O)$.
    \end{proof}
    
    Next we consider a particular kind of stream of edges, where each edge appears twice: given an arbitrary order of the vertices, for each vertex, all its incident edges are given in a row. For this \emph{incidence streaming model}~\cite{BravermanLSVY18} (sometimes called \emph{adjacency list model}~\cite{McGregorTV2021}), the next theorem shows that we can use much less space with just a single pass. 
    
    \begin{theorem}
        \label{thm:stream-simple}
        In the incidence streaming model, one can extract a $(1-\varepsilon)$-approximate solution using $O((\frac{k}{\varepsilon})^2)$ variables for uniform, partition, laminar, and $O((\frac{k}{\varepsilon} + k)^2)$ for transversal matroids.
    \end{theorem}
    
    \begin{proof}
        Let $\varepsilon > 0$. Given the type of the matroid $\mathcal{M}$, choose the corresponding value of $\tau$ as prescribed in Theorem~\ref{thm:greedyonUnion}. Start with an empty kernel $V' = \emptyset$. Through the execution of the algorithm, $V'$ will contain the largest independent set in $\tau\mathcal{M}$ with respect to the sum of the weighted degrees. When we process a vertex $v$ (\emph{i.e.} its set of incident edges) we can compute its weighted degree $\deg_w(v)$ and store the edges linking $v$ to elements of $V'$. If $V' + v$ is not independent in $\tau \mathcal{M}$, consider the  element with the smallest weighted degree $u$ in the circuit formed in $V' + v$. Then, set $V' \gets (V' + v) - u$. If $v$ is added into $V'$, we keep in memory all the edges linking $v$ to other vertices of $V'$. When an element is discarded or evicted from $V'$, all its incident edges are deleted. As a result, at any time during the execution of the algorithm, only $O(\tau^2)$ edges are stored, so the overall memory consumption is $O(\tau^2)$.
        
        In the end, we obtain exactly the approximate kernels described in the previous sections, and because we know all the values of the weighted degrees of $V'$ as well as the weights of the edges between them we can find the largest vertex cover in that kernel using brute force.
    \end{proof}
    
    \begin{remark}
    \label{rem:coverageStreaming}
        This model has an interesting interpretation 
        in the context of coverage function\footnote{A coverage function $f$ over a ground set $\{1,\dots,m\}$, associated with a universe $U$ of weighted elements and $m$ sets $A_1,\dots,A_m$, where $A_i \subseteq U$ for all $i$, is defined over all $S \subseteq \{1,\dots,m\}$ so that $f(S)$ is the sum of the weight of the elements in $\cup_{i \in S}A_i$. The frequency of an element of the universe is the number of sets $A_i$ it appears in. Here in our problem of maximum vertex cover, the vertices correspond to the ground set and the edges to the universe $U$. Note that for the special case of a vertex cover, the frequency (the maximum number of sets where an element of the universe appears in) is exactly 2, as an edge has only two endpoints.} maximization in the streaming setting---
        here the sets arrive over time 
        in such a way that the values of singletons $f(\{v\})$, for $v \in V$, are revealed one by one. 
        In case where a coverage function has bounded frequency larger than $2$, we also present in Appendix~\ref{app:stream-hypergraph} a streaming algorithm. 
    \end{remark}
    
\section{Local Search Algorithms for General Matroids} \label{sec:local-search}

    \subsection{A $2/3$ Polynomial-Time Local Search Approximation Algorithm}
        \label{sec:local-search-2/3}
        In this part we borrow ideas from~\cite{Filmus2012a}, using a \emph{non-oblivious} local search technique, meaning that instead of optimizing our objective function $E_G$ we use a potential function to guide the algorithm. Precisely, we define a potential function $g$:
        \begin{equation}g(S) = \sum_{e \in E} \alpha_{\#(e, S)}
        \label{equ:definitionG}
        w(e)\end{equation}
        where $\#(e, S)$ denotes the number of times the edge $e$ is covered by vertices of $S$ (hence, $\#(e, S) \in \{0,1,2\}$ in practice). The values $\alpha_0$, $\alpha_1$, and $\alpha_2$ have to be chosen carefully to optimize the approximation ratio. We suppose that $\alpha_2 - \alpha_1 \leq \alpha_1 - \alpha_0$, which is a quite intuitive assumption of decreasing marginal returns. Moreover, this guarantees that the function 
        $g$ is submodular (as in~\cite{Filmus2012a}). We set $\alpha_0 = 0$ and $\alpha_1 = 1$ and 
        will decide the value of $\alpha_2$ later. 
        
        Let $\varepsilon > 0$. The algorithm starts by a standard greedy step, which outputs a $1/2$ approximation $S$ of the maximum value of $g$ for a basis of $\mathcal{M}$. Then, we have a local search phase that tries to improve $S$ by exchanging an element of $s \in S$ with another element $s' \in V \backslash S$ if the following inequality holds:
        \[g(S - s + s') > (1 + \varepsilon)g(S),\]
        where the factor $(1 + \varepsilon)$ allows us to bound the number of improvement steps by $\log_{1 + \varepsilon}2$.
        
        \begin{algorithm}
    	\caption{Local search algorithm for maximum vertex cover under matroid constraint} 
    	\label{Alg:local-search}
    	\begin{algorithmic}[1]
    	\Function{Local-Search}{$\mathcal{M}, V, E, \varepsilon$}
        	\State $S\gets$ a basis in $\mathcal{M}$ obtained by the standard greedy algorithm for maximizing $g$
        	\While{$\exists (s,s') \in S \times V \backslash S$ such that $g(S - s + s') > (1 + \varepsilon)g(S)$ and $S - s + s' \in \mathcal{I}$}
        	    \State $S \gets S - s + s'$
        	\EndWhile
        	\State \Return $S$
    	\EndFunction
    	\end{algorithmic}
    	\end{algorithm}
        
        Suppose that the optimal base of our initial problem is $O$, and that our local search algorithm stopped at some solution $S$. Then there exists a bijection $\pi : S \rightarrow O$ such that for all $s \in S$ we have $S - s + \pi(s) \in \mathcal{I}$ and elements of $S \cap O$ are fixed points of $\pi$ (Corollary 39.12a in~\cite{Schrijver2003}). Hence, 
        \[\forall s \in S,\, (1 + \varepsilon) g(S) \geq g(S - s + \pi(s)),\]
        and therefore
        \[k(1 + \varepsilon)g(S) \geq \sum_{s \in S} g(S - s + \pi(s)).\]
        
        Now we define a partition of $E_{x, y, z}$ of $E$, where $x, y, z \in \{0,1,2\}$, as follows: the set $E_{x, y, z}$ contains the elements $e \in E$ if $e$ is covered by $x$ vertices in $S \backslash O$, by $y$ vertices in $S \cap O$, and by $z$ vertices in $O \backslash S$. By this definition of the partition of $E$, we can re-write the previous inequality as 
        \begin{equation}k(1+\varepsilon) \sum_{x, y, z}\alpha_{x + y} w(E_{x, y, z}) \geq \sum_{x, y, z} \beta_{x, y, z} w(E_{x, y, z}),
        \label{equ:XYZ}
        \end{equation}
        where the values of $\beta_{x, y, z}$ are:
        \begin{align*}
            \beta_{2,0,0} &= (k - 2) \alpha_2 + 2\\
            \beta_{0,2,0} &= k \alpha_2\\
            \beta_{0,0,2} &= 2 \\
            \beta_{1,0,1} &= \min\{(k-2) + \alpha_2, k\} = (k-2) + \alpha_2\\
            \beta_{1,1,0} &= (k - 1)\alpha_2 + 1\\
            \beta_{0,1,1} &= (k - 1) + \alpha_2\\
            \beta_{1,0,0} &= (k - 1)\\
            \beta_{0,1,0} &= k\\
            \beta_{0,0,1} &= 1
        \end{align*}
        The only tricky case is when $x = 1$, $y = 0$, and $z = 1$ because we do not know whether the two elements $s \in S$ and $o \in O$ covering $e$  satisfy $\pi(s) = o$ or $\pi(s) \neq o$, giving respectively coefficients of $k\alpha_1$ or $(k-2)\alpha_1 + \alpha_2 + \alpha_0$, and so we choose the smaller one, using the fact that $\alpha_2 - \alpha_1 \leq \alpha_1 - \alpha_0$.
        
        By adding $w(O)$ to both sides of~(\ref{equ:XYZ}) and re-arranging terms, we have  
        \begin{equation}\sum_{x, y, z}(k(1+\varepsilon) \alpha_{x + y} - \beta_{x, y, z} + \mathbbm{1}_{y + z \geq 1}) \cdot w(E_{x, y, z}) \geq w(O).
        \label{equ:wObothSides}
        \end{equation}
        We define the coefficients on the left-hand side as 
        \[\gamma_{x, y, z} = k(1+\varepsilon) \alpha_{x + y} - \beta_{x, y, z} + \mathbbm{1}_{y + z \geq 1}.\]
        If, for all $x + y \geq 1$ we have $\gamma_{x, y, z} \leq \theta$ and $\gamma_{x, y, z} \leq 0$ when $x = y = 0$, \\
        then $\theta \cdot w(S) \geq w(O)$.
        Now we write down all these coefficients:
        \begin{align*}
            \gamma_{2,0,0} &= (k + k\varepsilon) \alpha_2 - ((k - 2) \alpha_2 + 2) &= (2 + k\varepsilon) \alpha_2 - 2 \leq \theta\\
            \gamma_{0,2,0} &= (k + k\varepsilon) \alpha_2 - k \alpha_2 + 1 &= k\varepsilon \alpha_2 +  1 \leq \theta\\
            \gamma_{0,0,2} &=-2 + 1 &= -1 \leq 0\\
            \gamma_{1,0,1} &= (k + k\varepsilon) - ((k-2) + \alpha_2) + 1 &= 3 + k\varepsilon - \alpha_2\leq \theta\\
            \gamma_{1,1,0} &= (k + k\varepsilon) \alpha_2 - ((k - 1)\alpha_2 + 1) + 1 &= (1 + k \varepsilon)\alpha_2 \leq \theta\\
            \gamma_{0,1,1} &= (k + k\varepsilon) - ((k - 1) + \alpha_2) + 1 &= 2 + k \varepsilon - \alpha_2 \leq \theta\\
            \gamma_{1,0,0} &= (k + k\varepsilon) - (k - 1) &= 1 + k \varepsilon \leq \theta\\
            \gamma_{0,1,0} &= (k + k\varepsilon) - k + 1 &= 1 + k \varepsilon \leq \theta\\
            \gamma_{0,0,1} &= -1 + 1&= 0 \leq 0
        \end{align*}
        By setting $\alpha_2 = 3/2$ we get $\theta = 3/2 \cdot (1 + k \varepsilon)$. We have thus a $\frac{2}{3(1+k\varepsilon)}$ approximation. 
        To get a clean $\frac{2}{3}$ approximation, we can use the partial 
        enumeration technique of Filmus and Ward~\cite{Filmus2012a}. 
        For the sake of completeness, we provide the details here. 
        
        Algorithm~\ref{contracted-local-search} proceeds by guessing 
        the vertex $u \in O$ with the largest weighted degree and then 
        applying the previous algorithm on the subgraph $G_u=(V-u, E-\delta(u))$, with the contracted matroid $\mathcal{M}_u=
        (V-u, \mathcal{I}_u)$ as constraint, and the error term $\varepsilon=\frac{1}{(3k)^2}$. For an element $u \in V$ we set $\mathcal{M}_u = (V - u, \mathcal{I}_u)$ as the \emph{contracted matroid} of $\mathcal{M}$ by $u$ such that $\mathcal{I}_u = \{S \subseteq V - u : S + u \in \mathcal{I}\}$. Note that $\mathcal{M}_u$ is still a matroid.

        \begin{algorithm}
    	\caption{Algorithm providing an approximation guarantee of $2/3$}\label{contracted-local-search}
    	\begin{algorithmic}[1]
    	\Function{Contracted-matroid-search}{$\mathcal{M}, V, E$}
        	\State $S_{opt} \gets \emptyset$
        	\For{$v \in V$}
        	    \State $S_{cur} \gets v \cup \Call{Local-Search}{\mathcal{M}_v, V - v, E - \delta(v), 1/(3k)^2}$
        	    \If{$w(S_{cur}) > w(S_{opt})$}
        	        $S_{opt} \gets S_{cur}$
        	    \EndIf
        	\EndFor
        	\State \Return $S_{opt}$
    	\EndFunction
    	\end{algorithmic}
    	\end{algorithm}
    	
    	Let $S_u$ be the solution returned by the procedure 
    	$\Call{Local-Search}{\mathcal{M}_u, V - u, E - \delta(u), 1/(3k)^2}$ and $O_u = O - u$.  
        Then  
        $\frac{3}{2} \cdot (1 + \frac{1}{3k}) \cdot E_{G_u}(S_u) \geq E_{G_u}(O_u)$.
    	Adding $\deg_w(u)$ to both sides we obtain 
    	\[\frac{3}{2} \cdot \left(1 + \frac{1}{3k}\right) \cdot E_{G_u}(S_u) + \deg_w(u) \geq E_{G}(O).\]
    	
    	The left-hand side can be re-written 
    	$\frac{3}{2} \cdot (1 + \frac{1}{3k}) \cdot E_{G_u}(S_u) + \deg_w(u) = \frac{3}{2} \cdot E_{G}(S) + \frac{1}{2k} \cdot E_{G_u}(S_u) - \frac{1}{2} \cdot \deg_w(e) \leq \frac{3}{2}E_G(S)$, where the inequality holds because 
    	as $u$ is the vertex with the largest weighted 
    	degree in $O$, $E_{G_u }(S_u) \leq  E_{G}(O)  \leq k\cdot \deg_w(u)$. We have thus established that Algorithm~\ref{contracted-local-search} 
    	gives a $2/3$ approximation. 
    	
    	In terms of time complexity, 
        Algorithm~\ref{contracted-local-search} performs a total of $O\left(n\cdot k \cdot m\cdot \frac{\log 2}{\log \left(1 + (3k^2)^{-1}\right)}\right) = O(n m k^3)$ arithmetic operations and $O(n^2 k^3)$ oracle queries. To see this, observe that a local improvement step 
        in Algorithm~\ref{Alg:local-search} sees each edge at most twice, thus taking $O(k\cdot m)$ arithmetic operations and $O(k \cdot n)$ matroid oracle queries. Moreover, there are at most $\log_{(1 + \varepsilon)} 2$ such steps after the greedy step, which takes $O(k \cdot m)$ arithmetic operations and oracle queries.
    	
    	\begin{theorem}
    	    Algorithm~\ref{contracted-local-search} is a local search algorithm providing a $2/3$ approximation, and uses $O(nmk^3)$ arithmetic operations and $O(n^2k^3)$ matroid oracle queries.
    	\end{theorem}

    	\subsection{A $3/4$ FPT Approximation Algorithm} 
    	    \label{sec:local-search-3/4}
    	We show how to improve the approximation ratio to $3/4$, if one is allowed to use FPT time. The key insight here is that if the local optimal $S$ is totally disjoint from the optimal $O$, then in the preceding proof, we no longer need to be concerned with the coefficients $\gamma_{0,2,0}, \gamma_{1,1,0},\gamma_{0,1,1}, \gamma_{0,1,0}$ and their corresponding constraints.  
    	
    	We can slightly modify our algorithm as follows (formally  stated in Algorithm~\ref{local-search-3/4}): each time a local optimum is reached, we guess one of its vertices to be part of the optimal solution and proceed recursively. There can be at most $k!$ guesses. By setting $\alpha_2=5/3$ we can ensure that $\theta = 4/3 \cdot (1 + k \varepsilon)$. Again, using the partial enumeration technique allows us to get a clean $3/4$ approximation.
        
        \begin{algorithm}
    	\caption{Variant of the local search algorithm}\label{local-search-3/4}
    	\begin{algorithmic}[1]
    	\Function{Local-Search-$3/4$}{$\mathcal{M}, V, E, \varepsilon$}
        	$S\gets$ a basis in $\mathcal{M}$ obtained by the standard greedy algorithm for maximizing $g$
        	\While{$\exists (s,s') \in S \times V \backslash S$ such that $g(S - s + s') > (1 + \varepsilon)g(S)$ and $S - s + s' \in \mathcal{I}$}
        	    \State $S \gets S - s + s'$
        	\EndWhile
        	\State $S_{opt}\gets S$
        	\For{$s \in S$}
        	    \State $S_{cur} \gets \text{\Call{Local-Search-$3/4$}{$\mathcal{M}_s, V - s, E - \delta(s), \varepsilon$}} + s$
        	    \If{$w(S_{cur}) > w(S_{opt})$} $S_{opt} \gets S_{cur}$ \EndIf
        	\EndFor
        	\State \Return $S_{opt}$
    	\EndFunction
    	\end{algorithmic}
    	\end{algorithm}
    	
    \subsection{Allowing Exchanges of Size $p$ Does Not Improve the Ratio}
        \label{subsec:p-exch-2/3}
        One could think that exchanging $p>1$ elements at the same time instead of only one during an improvement step might allow us to get a better ratio than in  Section~\ref{sec:local-search-2/3}. We show in this section that it is not possible, for any choice of $\alpha_2$.
        
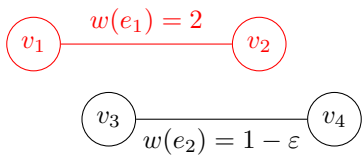
\begin{figure}[h]
            \centering
            \begin{tikzpicture}
                \node[draw, circle, red] (V1) at (0,0) {$v_1$};
                \node[draw, circle, red] (V2) at (3,0) {$v_2$};
                \node[draw, circle] (V3) at (1,-1) {$v_3$};
                \node[draw, circle] (V4) at (4,-1) {$v_4$};
                
                \draw[red] (V1) -- (V2) node[pos=0.5, above]{$w(e_1) = 2$};;
                \draw (V3) -- (V4) node[pos=0.5, below]{$w(e_2) = 1 - \varepsilon$};;
            \end{tikzpicture}
            \caption{Example when $\alpha_2 \geq 3/2$.}
            \label{fig:more-3/2}
        \end{figure}

        When $\alpha_2 \geq 3/2$, let $\varepsilon > 0$ and consider the example in Figure~\ref{fig:more-3/2}, where the constraint is imposed by a uniform matroid of rank $2$. In fact, the subset $S \subseteq V$ of size $2$ which optimizes $g$ is $\{v_1, v_2\}$, as $g(\{v_1, v_2\}) \geq 3$, but the optimal solution for the maximum cover problem is $\{v_1, v_3\}$, giving a value $g(\{v_1, v_3\}) = 3 - \varepsilon$. Therefore the solution given by a local search algorithm using $\alpha_2 \geq 3/2$ cannot have a better approximation guarantee than $2/3$.
        
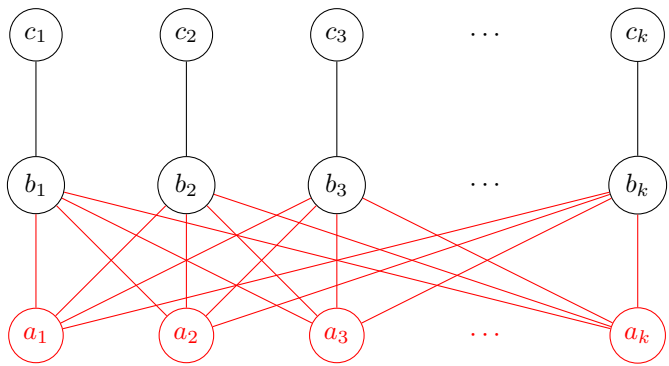
\begin{figure}[h]
            \centering
            \begin{tikzpicture}
                \node[draw, circle, red] (A1) at (0,0) {$a_1$};
                \node[draw, circle, red] (A2) at (2,0) {$a_2$};
                \node[draw, circle, red] (A3) at (4,0) {$a_3$};
                \node[red] (Ap) at (6,0) {$\cdots$};
                \node[draw, circle, red] (Ak) at (8,0) {$a_k$};
                
                \node[draw, circle] (B1) at (0,2) {$b_1$};
                \node[draw, circle] (B2) at (2,2) {$b_2$};
                \node[draw, circle] (B3) at (4,2) {$b_3$};
                \node (Bp) at (6,2) {$\cdots$};
                \node[draw, circle] (Bk) at (8,2) {$b_k$};
                
                \node[draw, circle] (C1) at (0,4) {$c_1$};
                \node[draw, circle] (C2) at (2,4) {$c_2$};
                \node[draw, circle] (C3) at (4,4) {$c_3$};
                \node (Cp) at (6,4) {$\cdots$};
                \node[draw, circle] (Ck) at (8,4) {$c_k$};
                
                \draw[red] (A1) -- (B1);
                \draw[red] (A1) -- (B2);
                \draw[red] (A1) -- (B3);
                \draw[red] (A1) -- (Bk);
                
                \draw[red] (A2) -- (B1);
                \draw[red] (A2) -- (B2);
                \draw[red] (A2) -- (B3);
                \draw[red] (A2) -- (Bk);
                
                \draw[red] (A3) -- (B1);
                \draw[red] (A3) -- (B2);
                \draw[red] (A3) -- (B3);
                \draw[red] (A3) -- (Bk);
                
                \draw[red] (Ak) -- (B1);
                \draw[red] (Ak) -- (B2);
                \draw[red] (Ak) -- (B3);
                \draw[red] (Ak) -- (Bk);
                
                \draw (B1) -- (C1);
                \draw (B2) -- (C2);
                \draw (B3) -- (C3);
                \draw (Bk) -- (Ck);
                
            \end{tikzpicture}
            \caption{Example when $\alpha_2 < 3/2$. The edges between the $a_i$s and the $b_i$s are of weight $1$, and those between the $b_i$s and the $c_i$s are of weight $k/2$.}
            \label{fig:less-3/2}
        \end{figure}

        When $\alpha_2 < 3/2$, we use the example depicted in Figure~\ref{fig:less-3/2}, where the constraint is a uniform matroid of rank $k$. We show that for any $p$, if $k$ is large enough, the set $S = \{a_i\}_{1 \leq i \leq k}$ is a local optimum of $g$, even if $p$ simultaneous exchanges at the same time are allowed. As the solution $S$ has value $E_G(S) = k^2$ while the optimal solution $O = \{b_i\}_{1 \leq i \leq k}$ has value $E_G(O) = k^2 + k^2/2$, the approximation guarantee of such an algorithm cannot be better than $2/3$. 
        
        We can observe that the most interesting exchanges have to be of the form $a_i \leftrightarrow b_i$. If we perform $p$ such exchanges, for instance $a_1, \dots, a_p \leftrightarrow b_1, \dots, b_p$ then the gain for the value of $g$ is:
        \begin{align*}
            &\, g(S + \{b_1, \dots b_p\} - \{a_1, \dots a_p\}) - g(S)\\
            &= p(k-p)(\alpha_2 - 1) &\text{for $\{a_i,b_i\}$ such that $i > p,\,j\leq p$}\\
            &\quad - p(k-p) &\text{for $\{a_i,b_i\}$ such that $i \leq p,\,j > p$}\\
            &\quad + p k/2 &\text{for $\{b_i,c_i\}$ such that $i \leq p$}\\
            &= p (k-p)(\alpha_2 - 2) + p k/2\\
            &= p k(\alpha_2 - 3/2) + p^2(2 - \alpha_2)
        \end{align*}
        as $\alpha_2 < 3/2$, when $k$ is big enough the gain becomes negative. Therefore, with a large $k$, $S = \{a_i\}_{1 \leq i \leq k}$ is a local optimum even if $p$ exchanges are allowed. 
        
        \paragraph*{What if the degrees are bounded?} Our previous example uses vertices of large degrees, so one is naturally led to ask what happens if the vertex degree is bounded. Could one obtain approximation ratios significantly better than $2/3$ in this case? In the following, we can assume that $\alpha_2 < 3/2$ (otherwise the first example in Figure~\ref{fig:more-3/2} shows the approximation ratio cannot be better than $2/3$).

        Let $d \geq 2$ be the desired bound for the degrees in our graph. We proceed somehow similarly to Lee et al.~\cite{Lee2010a}. Using the result of~\cite{lazebnik1995new}, we know there exists  a bipartite, $d$-regular graph $G = (V, E)$ of girth at least $2p + 1$ (\emph{i.e.} the smallest circuit in the graph is of length at least $2p+1$).\footnote{In~\cite{lazebnik1995new} the construction is specific for a prime power $d'$, but we can use their construction to obtain our $d$-regular bipartite graph for some $d \leq d'$ easily: by removing iteratively a perfect matching from the graph. Such a perfect matching exists because of Hall's marriage theorem, \emph{e.g.},~\cite{Schrijver2003}.} Let $A$ and $B$ be the bi-partition of the graph, so that $V = A \cup B$. As it is a bipartite regular graph, there exists a perfect matching $M$ in that graph. We denote $\{a_1, b_1\},\dots, \{a_k, b_k\}$ such a matching. We build a graph $H = (V', E')$ as  follows: we take the vertices and the edges of $G$, we add a set of vertices $C = \{c_1, \dots, c_k\}$, then we add edges $\{b_i,c_i\}$ for $1 \leq i \leq k$ and we delete the edges of the matching $\{a_i, b_i\}$ for $1 \leq i \leq k$. Clearly in $H$ the degree bound is still $d$. We define the weights $w(\{a_i, b_j\}) = 1$ for $1 \leq i, j \leq k$ and $w(\{b_i, c_i\}) = \lambda$ (we will set the value of $\lambda$ later).
        
        We define a partition matroid $\mathcal{M} = (V', \mathcal{I})$ with the partition $V_1 = \{a_1,b_1,c_1\},\dots,V_k=\{a_k,b_k,c_k\}$ and a bound of one element per set $V_i$. 
        
        We want to show that $A$ is a local optimum, whereas the real optimum is $B$. In fact, $E_{H}(A) = k(d-1)$ and $E_{H}(B) = k(\lambda + d - 1)$. The only relevant exchanges are of the form $a_i \leftrightarrow b_i$, as taking $c_i$ is clearly less interesting. Consider the indexes $i_1, \dots, i_p$ of the exchanges tried during a local search step.
        \begin{align*}
            &\, g(A + \{b_{i_1},\dots, b_{i_p}\} - \{a_{i_1},\dots, a_{i_p}\}) - g(A)\\
            &= - (p(d - 1) - s) &\text{edges in $\{a_{i_1},\dots, a_{i_p}\} \times B\backslash \{b_{i_1},\dots, b_{i_p}\}$}\\
            &\quad + p \cdot \lambda &\text{for $\{b_{i_j}, c_{i_j}\}$, for $1 \leq j \leq p$}\\
            &\quad + (\alpha_2 - 1) (p (d - 1) - s) &\text{edges in $\{b_{i_1},\dots, b_{i_p}\} \times A\backslash \{a_{i_1},\dots, a_{i_p}\}$}\\
            &= (\alpha_2 - 2)(p(d-1) - s) + p \cdot \lambda\\
            &\leq (\alpha_2 - 2) (p(d-2) + 1) + p \cdot \lambda & \text{As $s \leq p-1$ and $\alpha < 2/3$}
        \end{align*}
        where $s$ denote the number of ``saved'' edges. 
        Precisely, these are the edges between 
        $\{a_{i_1},\dots, a_{i_p}\}$ and 
        $\{b_{i_1},\dots, b_{i_p}\}$ in the graph $H$. 
        We call them ``saved'' because they are still covered after the exchange, unlike other edges incident to the $a_{i_j}$s. We now claim that $s\leq p-1$. Recall that as the original graph $G$ has girth at least $2p+1$, there cannot be a cycle among the vertices in $\{a_{i_1},\dots, a_{i_p}\} \cup \{b_{i_1},\dots, b_{i_p}\}$. So the subgraph of $G$ induced by these vertices forms a forest, with at most $2p-1$ edges. Recall that when constructing $H$, we have removed the edges $\{a_i,b_i\}$ for all $i$. Therefore $p$ edges in the subgraph are gone. We conclude that at most $p-1$ edges of that forest can be the saved edges. 
        
        Setting $\lambda = (2 - \alpha_2) \left((d-2) + 1/p\right)$, the gain becomes non-positive, and $E_{H}(B) = k((2 - \alpha_2) \left((d-2) + 1/p\right) + d - 1) = (3 - \alpha_2)E_{H}(A) - k(2 - \alpha_2)(1 - 1/p)$, so 
        \[\frac{E_{H}(B)}{E_{H}(A)} = (3 - \alpha_2) - \frac{(2 - \alpha_2)(1 - 1/p)}{d-1}.\]
        This implies that the approximation ratio is bounded by $2/3+\theta(1/d)$, as $\alpha_2 \leq 3/2$. 

\section{Local Search Algorithm for Two Matroids}

    In this section, we assume that two matroids $\mathcal{M}_1 = (V, \mathcal{I}_1)$ and $\mathcal{M}_2 = (V, \mathcal{I}_2)$ are imposed on the vertex set $V$. 
    
    The idea of the \emph{exchange graph}~\cite[Section 41.2]{Schrijver2003}, is defined as follows. 
    Given $I \in \mathcal{I}_1 \cap \mathcal{I}_2$, an oriented (bipartite) digraph $D_{\mathcal{M}_1,\mathcal{M}_2}(I)$ has the 
    following two types of edges: 
    \begin{itemize}
        \item for $i \in I$ and $j \in V \backslash I$ there is an arc $(i, j)$ if $I - i + j \in \mathcal{I}_1$
        \item for $i \in I$ and $j \in V \backslash I$ there is an arc $(j, i)$ if $I - i + j \in \mathcal{I}_2$
    \end{itemize}
    In this exchange graph, a dipath/dicycle $A$ (whose vertices are denoted $V(A)$) is called \emph{feasible} if $I\Delta V(A)$ is in $\mathcal{I}_1 \cap \mathcal{I}_2$ and if for any sub-dipath $A'$ of $A$ such that the endpoints of $A'$ are in $I$ (or are the endpoints of $A$) then $I\Delta V(A')$ is also in $\mathcal{I}_1 \cap \mathcal{I}_2$.
    
    The following exchange lemma from~\cite{Lee2010a} will be extremely useful for our purpose.
    \begin{lemma}[Lemma~2.5 in~\cite{Lee2010a}]\label{lem:exchange}
        Let $I, J \in \mathcal{I}_1 \cap \mathcal{I}_2$. Then there exists an integer $s \geq 0$ and a collection of dipaths/dicycles $\{D_1, \dots, D_t\}$ (possibly with repetitions), feasible in the digraph $\mathcal{D}_{\mathcal{M}_1, \mathcal{M}_2}(I)$, using only elements of $I \Delta J$, so that each element of $I \Delta J$ appears in exactly $2^s$ dipaths/dicycles $D_i$.
    \end{lemma}
    
	Suppose that we have a solution $S \in \mathcal{I}_1 \cap \mathcal{I}_2$. Let $O \in \mathcal{I}_1 \cap \mathcal{I}_2$ be the global optimum. By Lemma~\ref{lem:exchange} with $I = S$ and $J = O$, there exists $s \geq 0$ and a collection of dipaths/dicycles $\{D_1, \dots, D_t\}$ such that each element of $S \Delta O$ appears exactly $2^s$ times in these dipaths/dicycles. We could potentially perform local search exchanges corresponding to these dipaths and dicycles to improve our solution $S$. However, such paths and cycles can be arbitrarily long, thus implying a prohibitively high running time. To overcome this issue, we will use an idea of Lee et al.~\cite[Lemma 3.1]{Lee2010a}, in which these long paths and cycles are cut into pieces of length $O(p)$ for some constant $p$. Such pieces will then allow us to perform local search in polynomial time. 
    
    Precisely, for each dipath/dicycle $D_i$, we number the vertices of $O \backslash S$ along the path consecutively. For each $D_i$, we create $p + 1$ copies $D_{i, 0}, \dots, D_{i,p}$ and in the copy $D_{i, q}$ we throw away vertices that are labeled $q$ modulo $p+1$. By cutting these dipaths/dicycles, now the corresponding exchanges contain only up to $2p$ vertices from $O \backslash S$ and at most $2p + 1$ vertices from $S \backslash O$. Notice that these shortened paths $A$ remain feasible, i.e., $I\Delta V(A) \in \mathcal{I}_1 \cap \mathcal{I}_2$. 
    
    \begin{algorithm}
	\caption{Local search algorithm for maximum vertex cover under two matroid constraints} \label{local-search-2matroid}
	\begin{algorithmic}[1]
	\Function{Local-Search-2Matroids}{$\mathcal{M}, V, E, p, \varepsilon$}
    	\State $S\gets$ a common independent set in  $\mathcal{I}_1 \cap \mathcal{I}_2$ obtained in polynomial time and providing a constant approximation ratio (for instance the algorithm in~\cite{Lee2010a}, or just greedy)
    	\While{$\exists A \subseteq S, B \subseteq V \backslash S$ such that $g(S - A + B) > (1 + \varepsilon)g(S)$, $|A| \leq 2p + 1$, $|B| \leq 2p$, and $S - A + B \in \mathcal{I}_1 \cap \mathcal{I}_2$}
    	    \State $S \gets S - A + B$
    	\EndWhile
    	\State \Return $S$
	\EndFunction
	\end{algorithmic}
	\end{algorithm}
	
	Algorithm~\ref{local-search-2matroid}, a local search algorithm allowing exchanges size up to $2p$, exploits exactly the above idea. 
	Here the potential function $g$ is defined by~(\ref{equ:definitionG})
	as in Section~\ref{sec:local-search-2/3}. Again $\alpha_0=0$, $\alpha_1=1$, and we decide the value of $\alpha_2$ later. 
	
	Consider the family $\{D_{i, q}\}_{1 \leq i \leq t, 0 \leq q \leq p}$ and all the exchanges they contain,
	which are denoted $\{(A_i, B_i) \subseteq S \backslash O \times O \backslash S\}_{1 \leq i \leq t'}$. Each element of $O \backslash S$ appears in exactly $p 2^s$ exchanges and each element of $S \backslash O$ in $(p+1) 2^s$ exchanges. As $S$ is a local optimum, all these exchanges satisfy the inequality of Algorithm~\ref{local-search-2matroid}:
	\[\forall 1 \leq i \leq t', (1 + \varepsilon) g(S) \geq g(S - A_i + B_i),\]
	and summing over all these inequalities:
	\[t'(1 + \varepsilon) g(S) \geq \sum_{i = 1}^{t'}g(S - A_i + B_i).\]
	Using the sets $E_{x, y, z}$ as defined in Section~\ref{sec:local-search}, we have the inequality analogous to inequality~(\ref{equ:XYZ}):
	\[t'(1 + \varepsilon) g(S) = t' (1 + \varepsilon) \sum_{x, y, z}\alpha_{x+y} w(E_{x, y, z}) \geq \sum_{i = 1}^{t'}g(S - A_i + B_i) \geq \sum_{x, y, z}\beta'_{x,y,z} w(E_{x, y, z})\]
	where the $\beta'_{x, y, z}$ are equal to:
	\begin{align*}
	    \beta'_{2,0,0} &= (t' - (p+1)2^s) \alpha_2\\
	    \beta'_{0,2,0} &= t'\alpha_2\\
	    \beta'_{0,0,2} &= p2^s\alpha_2\\
	    \beta'_{1,0,1} &= t' - (2p+1)2^s + p2^s\alpha_2\\
	    \beta'_{1,1,0} &= (t' - (p+1)2^s)\alpha_2 + (p+1)2^s\\
	    \beta'_{0,1,1} &= t' - p2^s + p2^s\alpha_2\\
	    \beta'_{1,0,0} &= t' - (p + 1)2^s\\
	    \beta'_{0,1,0} &= t'\\
	    \beta'_{0,0,1} &= p2^s
	\end{align*}
	
	There are only two nontrivial cases: when $(x,y,z) = (2,0,0)$, the worst situation is when, for a given edge $e$, all the exchanges involve both of its two endpoints. To see this, recall that the two endpoints appear in total of $2(p+1)2^s$ times in the family $\{D_i,q\}_{1\leq i \leq t, 0 \leq q \leq p}$, so if $l$ exchanges involve both endpoints, the coefficient for this edge is $(t' - 2(p+1)2^s + l)\alpha_2 + (2(p+1)2^s - 2l)$, which is minimized when $l = p2^s$. For the other case $(x,y,z) = (0,0,2)$, the worst situation for an edge $e$ is similarly when all the exchanges involve both of its endpoints. 
	
    Similar to the previous section, we can derive an inequality analogous to (\ref{equ:wObothSides}):
    \[\sum_{x, y, z}(k(1+\varepsilon) \alpha_{x + y} - \beta'_{x, y, z} + p2^s \cdot \mathbbm{1}_{y + z \geq 1}) \cdot w(E_{x, y, z}) \geq p2^s \cdot w(O).\]
    
    Similar to the previous section, we define 
    \[\gamma'_{x, y, z} = k(1+\varepsilon) \alpha_{x + y} - \beta_{x, y, z} + p2^s \cdot \mathbbm{1}_{y + z \geq 1}.\]

    If for all $x + y + z \geq 2$ we have $\gamma'_{x, y, z} \leq \theta$, and if $x = y = 0$ we have $r'_{x,y,z} \leq 0$,  
    then $\theta \cdot w(S) \geq p2^s \cdot w(O)$.
    We write down all these coefficients and their corresponding constraints:
    \begin{align*}
	    \gamma'_{2,0,0} &= (t' + t'\varepsilon)\alpha_2 - (t' - (p+1)2^s) \alpha_2 &= (p+1)2^s\alpha_2 + t'\varepsilon \alpha_2 \leq \theta\\
	    \gamma'_{0,2,0} &= (t' + t'\varepsilon)\alpha_2 - t'\alpha_2 + p2^s &= p2^s + t'\varepsilon \alpha_2 \leq \theta\\
	    \gamma'_{0,0,2} &= - p2^s\alpha_2 + p2^s &= - p2^s\alpha_2 + p2^s \leq 0\\
	    \gamma'_{1,0,1} &= (t' + t'\varepsilon) - (t' - (2p+1)2^s + p2^s\alpha_2) + p2^s &= (3p + 1)2^s - p2^s\alpha_2 + t'\varepsilon \leq \theta\\
	    \gamma'_{1,1,0} &= (t' + t'\varepsilon)\alpha_2 \\ &\quad - ((t' - (p+1)2^s)\alpha_2 + (p+1)2^s) + p2^s &= (p+1)2^s\alpha_2 - 2^s + t'\varepsilon\alpha_2 \leq \theta\\
	    \gamma'_{0,1,1} &= (t' + t'\varepsilon) - (t' - p2^s + p2^s\alpha_2) + p2^s &= 2p2^s - p2^s\alpha_2 + t'\varepsilon \leq \theta\\
	    \gamma'_{1,0,0} &= (t' + t'\varepsilon) - (t' - (p + 1)2^s) &= (p + 1)2^s + t'\varepsilon \leq \theta\\
	    \gamma'_{0,1,0} &= (t' + t'\varepsilon) - t' + p2^s &= p2^s + t'\varepsilon \leq \theta\\
	    \gamma'_{0,0,1} &= - p2^s + p2^s &= 0 \leq 0
	\end{align*}
	setting $\alpha_2 = 3/2$ we get $\theta = (p + 1)2^s \cdot 3/2 + t' \varepsilon \cdot 3/2$, and therefore
	\[\left(\frac{3}{2}\left(1 + \frac{1}{p}\right) + \frac{3t'\varepsilon}{2p2^s}\right) \cdot w(S) \geq w(O).\]
	If we denote $k = \min\{k_1,k_2\}$ the minimum rank among the two matroids, then we know that $t' \leq k 2^s$, therefore we get:
    \[\left(\frac{3}{2}\left(1 + \frac{1}{p}\right) + \frac{3k\varepsilon}{2p}\right) \cdot w(S) \geq w(O).\]
    Using the same partial enumeration technique as in Section~\ref{sec:local-search}, one can get, for a fixed $p$, a polynomial-time algorithm providing a $2/3 \cdot (1 - 1/(p+1))$ approximation.
    
    \begin{theorem}
        Algorithm~\ref{local-search-2matroid} is a local search algorithm providing a $2/3 \cdot (1 - 1/(p+1))$ approximation using $n^{O(p)}$ arithmetic operations and matroid oracle calls. 
    \end{theorem}
    
\section{Conclusion and Open Questions}

    Theorem~\ref{thm:greedyonUnion} allows us to generalize the cardinality constraint to some special cases of matroid constraints, and these ideas could be useful for other kernelization algorithms. Regarding the bounds of Theorem~\ref{thm:greedyonUnion}, tight examples can be built to show that the values of $\tau$ provided are the best possible for partition and laminar matroids. For transversal matroids it is less clear whether the bound for $\tau$ can be improved or not. The most important open question regarding this result is whether Theorem~\ref{thm:greedyonUnion} can be generalized to other types of matroids.
    
    Regarding local search, we have shown some limitations of the current techniques, which cannot exceed a $2/3$ approximation. Thus to close the gap between local search and LP techniques one would have to use new techniques, for instance by introducing a new type of objective function for this problem.

\bibliography{library}

\appendix

\section{An FPT-AS for Hypergraphs}
    \label{app:hypergraph}
    Suppose that we are given a hypergraph $G=(V,E)$ with edge size bounded by a constant $\eta \geq 2$. We proceed as in Section~\ref{sec:kernelization}. First, notice that:
    \[E_G(S) = E_G(O^{in}) + E_G(U^*) - E_G(O^{in}, U^*).\]
    We bound $\mathbb{E}[E_G(O^{in}, U^*)]$ as follows. By construction, $\mathbb{P}[u \in U^*] = \varepsilon$ for all $u \in U$. Then, 
    \begin{align*}
        \mathbb{E}[E_G(O^{in}, U^*)] = \sum_{e \in E, e \cap O^{in} \neq \emptyset} w(e) \cdot \mathbbm{1}[e \cap U^* \neq \emptyset] &\leq \sum_{e \in E, e \cap O^{in} \neq \emptyset} w(e) \cdot (\eta - 1) \cdot \varepsilon \\&= \varepsilon \cdot (\eta - 1) \cdot \mathbb{E}[E_G(O^{in})].
    \end{align*}
    using union bound and the fact that at most $\eta - 1$ endpoints can be in $U$.
    Furthermore, the value $\mathbb{E}[E_G(U^*)]$ can be rearranged as follows: 
    \begin{align*}
        &\mathbb{E}[E_G(U^*)] = \mathbb{E}\left[\sum_{u \in U^*}\left(\deg_w(u) - \sum_{e \in \delta(u)}w(e) \cdot \frac{|e \cap U^*| - 1}{|e \cap U^*|}\right)\right]\\
        \geq&\, \mathbb{E}\left[\sum_{u \in U^*}\left(\deg_w(u) - \sum_{e \in \delta(u)}w(e) \cdot \frac{\eta - 1}{\eta} \cdot \mathbbm{1}[e \cap U^* \backslash\{u\} \neq \emptyset]\right)\right]\\
        \geq&\, \mathbb{E}\left[\sum_{u \in U}\left(\deg_w(u) \cdot \mathbbm{1}[u \in U^*]- \sum_{e \in \delta(u)}w(e) \cdot \frac{\eta - 1}{\eta} \cdot \mathbbm{1}[u \in U^* \wedge e \cap U^* \backslash\{u\} \neq \emptyset]\right)\right]\\
        \geq&\, \mathbb{E}\left[\sum_{u \in U}\left(\deg_w(u) \cdot \varepsilon - \frac{\eta - 1}{\eta} \sum_{e \in \delta(u)}w(e) \cdot (\eta - 1) \cdot \varepsilon ^2\right)\right]\\
        &\geq \varepsilon \cdot (1 - \varepsilon \cdot (\eta - 1)) \left(\sum_{u \in U} \deg_w(u)\right).
    \end{align*}
    
    Recall that by robustness, for all $u \in O^{out}$, the elements of $U_{u}$ have weighted degree no less than the one of $u$. Therefore, 
    \begin{align*}
        \mathbb{E}[E_G(U^*)] &\geq \varepsilon \cdot (1 - \varepsilon \cdot (\eta - 1)) \left(\sum_{u \in O^{out}}\sum_{v \in U_{u}}\deg_w(v)\right)\\
        &\geq \varepsilon \cdot (1 - \varepsilon \cdot (\eta - 1)) \left(\sum_{u \in O^{out}}\frac{1}{\varepsilon} \cdot \deg_w(u)\right)\\
        &\geq (1 - \varepsilon \cdot (\eta - 1)) \cdot E_G(O^{out}).
    \end{align*}

    As a result, we get:
    \begin{align*}
        \mathbb{E}[E_G(S)] &\geq E_G(O^{in}) + (1 - \varepsilon\cdot(\eta - 1)) \cdot E_G(O^{out}) - \varepsilon \cdot (\eta - 1) \cdot E_G(O^{in})\\
        &\geq (1 - \varepsilon \cdot (\eta - 1)) \cdot E_G(O).
    \end{align*}
    By averaging principle, there exists a set $S \subseteq V'$ such that $S \in \mathcal{I}$ and such that $E_G(S) \geq (1 - (\eta - 1) \cdot \varepsilon) \cdot E_G(O)$.
    
\section{Streaming FPT-AS Algorithm for Hypergraphs}
    \label{app:stream-hypergraph}
    Here we suppose that we are given a hypergraph $G=(V,E)$ with edge size bounded by a constant $\eta \geq 2$ as an adjacency list stream. Using the idea of Theorem~\ref{thm:stream-simple} and the result of Appendix~\ref{app:hypergraph}, one can get in the incidence streaming model a $(1-(\eta-1) \cdot \varepsilon)$ approximation using $O(\tau^\eta)$ memory, where $\tau$ depends on the type of matroid, as prescribed in Theorem~\ref{thm:greedyonUnion}. In fact, we can maintain through the execution of the algorithm for each subset of at most $\eta$ elements $S \subseteq V'$ the sum of the weight of the hyper-edges $e$ such that $e \cap V' = S$ (just like in the proof of Theorem~\ref{thm:stream-simple} where we keep track of these values for pairs in $V'$). For instance, for a uniform, a partition, or a laminar matroid, we could get a $(1 - \varepsilon)$ approximation using $O((\frac{2\eta \cdot k}{\varepsilon})^\eta)$ variables. This shows that for the special matroids that we studied of rank $k$, a weighted coverage function with bounded frequency $\eta$ can be $(1 - \varepsilon)$ approximated in streaming. This extends a result of~\cite{McGregorTV2021} to matroids.
    
\section{Local Search with the Objective Function $E_G$}
    \label{app:oblivious}
\begin{figure}[h]
        \centering
        \begin{tikzpicture}
            \node[draw, circle, red] (A1) at (0,0) {$a_1$};
            \node[draw, circle, red] (A2) at (2,0) {$a_2$};
            \node[draw, circle, red] (A3) at (4,0) {$a_3$};
            \node[red] (Ap) at (6,0) {$\cdots$};
            \node[draw, circle, red] (Ak) at (8,0) {$a_k$};
            
            \node[draw, circle] (B1) at (0,2) {$b_1$};
            \node[draw, circle] (B2) at (2,2) {$b_2$};
            \node[draw, circle] (B3) at (4,2) {$b_3$};
            \node (Cp) at (6,2) {$\cdots$};
            \node[draw, circle] (Bk) at (8,2) {$b_k$};
            
            \node[draw, circle] (C1) at (0,4) {$c_1$};
            \node[draw, circle] (C2) at (2,4) {$c_2$};
            \node[draw, circle] (C3) at (4,4) {$c_3$};
            \node (Cp) at (6,4) {$\cdots$};
            \node[draw, circle] (Ck) at (8,4) {$c_k$};
            
            \draw[red] (A1) -- (B1);
            \draw[red] (A1) -- (B2);
            \draw[red] (A1) -- (B3);
            \draw[red] (A1) -- (Bk);
            
            \draw[red] (A2) -- (B1);
            \draw[red] (A2) -- (B2);
            \draw[red] (A2) -- (B3);
            \draw[red] (A2) -- (Bk);
            
            \draw[red] (A3) -- (B1);
            \draw[red] (A3) -- (B2);
            \draw[red] (A3) -- (B3);
            \draw[red] (A3) -- (Bk);
            
            \draw[red] (Ak) -- (B1);
            \draw[red] (Ak) -- (B2);
            \draw[red] (Ak) -- (B3);
            \draw[red] (Ak) -- (Bk);
            
            \draw (B1) -- (C1);
            \draw (B2) -- (C2);
            \draw (B3) -- (C3);
            \draw (Bk) -- (Ck);
            
        \end{tikzpicture}
        \caption{Example when the function used for local search is $E_G$. The edges between the $a_i$s and the $b_i$s are of weight $1$, and those between the $b_i$s and the $c_i$s are of weight $k\cdot (1 - \varepsilon)$.}
        \label{fig:less-1/2}
    \end{figure}
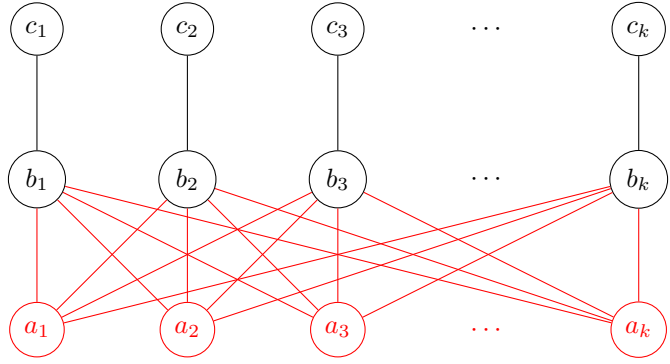

    Let $\varepsilon > 0$. When the cover function $E_G$ is used for the local search, we use the example depicted in Figure~\ref{fig:less-1/2}, where the constraint is a uniform matroid of rank $k$. This example is very similar to the one in Section~\ref{subsec:p-exch-2/3}. We show that for any $p$, if $k$ is large enough, the set $S = \{a_i\}_{1 \leq i \leq k}$ is a local optimum of $E_G$, even if $p$ simultaneous exchanges at the same time are allowed. As the solution $S$ has value $E_G(S) = k^2$ while the optimal solution $O = \{b_i\}_{1 \leq i \leq k}$ has value $E_G(O) = k^2 + k^2\cdot (1 - \varepsilon)$, the approximation guarantee of such an algorithm cannot be better than $1/2$. 
        
    We can observe that the most interesting exchanges have to be of the form $a_i \leftrightarrow b_i$. If we perform $p$ such exchanges, for instance $a_1, \dots, a_p \leftrightarrow b_1, \dots, b_p$ then the gain for the value of $g$ is:
    \begin{align*}
        &\quad E_G(S + \{b_1, \dots b_p\} - \{a_1, \dots a_p\}) - E_G(S)\\
        &= - p(k-p) &\text{for $\{a_i,b_i\}$ such that $i \leq p,\,j > p$}\\
        &\quad + p k\cdot (1 - \varepsilon) &\text{for $\{b_i,c_i\}$ such that $i \leq p$}\\
        &= - p (k-p) + p k \cdot (1- \varepsilon) = - p k \varepsilon + p^2
    \end{align*}
    as $\varepsilon > 0$, when $k$ is big enough the gain becomes negative. Therefore, when $k$ is big enough, $S = \{a_i\}_{1 \leq i \leq k}$ is a local optimum even if $p$ exchanges are allowed.
    
\end{document}